\documentclass{ws-ijmpe}
\usepackage{graphicx}
\pdfoutput=1
\makeatletter
\def\makerobust#1{%
  \@ifundefined{fr@gile\expandafter\@gobble\string#1}%
  {\expandafter
    \let\csname fr@gile\expandafter\@gobble\string#1\endcsname#1%
   \edef#1{\noexpand\protect\expandafter\noexpand
               \csname fr@gile\expandafter\@gobble\string#1\endcsname}}%
  {\@warning{\string#1 is already robust (command ignored)}}}
\makerobust{\cite}  
\makerobust{\refcite}
\makeatother

\begin{document}
\markboth{Rudolph C. Hwa}{Hadron Correlations in Jets and Ridges through Parton Recombination}

\catchline{}{}{}{}{}

\title{\bf HADRON CORRELATIONS IN JETS AND RIDGES THROUGH PARTON RECOMBINATION}

\author{\footnotesize RUDOLPH C. HWA}

\address{Institute of Theoretical Science and Department of
Physics\\ University of Oregon, Eugene, OR 97403-5203, USA\\
hwa@uoregon.edu}

\maketitle

\begin{abstract}
Hadron correlations in jets, ridges and opposite dijets at all $p_T$ above 2 GeV/c are discussed.  Since abundant data are available from RHIC at intermediate $p_T$, a reliable hadronization scheme at that $p_T$ range is necessary in order to relate the semihard partonic processes to the observables.  The recombination model is therefore first reviewed for that purpose.  Final-state interaction is shown to be important for the Cronin effect, large B/M ratio and forward production.  The effect of semihard partons on the medium is then discussed with particular emphasis on the formation of ridge with or without trigger.  Azimuthal anisotropy can result from ridges without early thermalization.  Dynamical path length distribution is derived for any centrality.  Dihadron correlations in jets on the same or opposite side are shown to reveal detail properties of trigger and antitrigger biases with the inference that tangential jets dominate the dijets accessible to observation.
\end{abstract}
\newpage
\tableofcontents
\markboth{Rudolph C. Hwa}{Hadron Correlations in Jets and Ridges through Parton Recombination}

\newpage
\section{Introduction}

Among the many properties of the dense medium that have been studied at RHIC, the nature of jet-medium interaction has become the subject of particular current interest.\cite{1.0,1.01}  Jet quenching, proposed as a means to reveal the effect of the hot medium produced in heavy-ion collisions on the hard parton traversing that medium,\cite{1.1,1.2} has been confirmed by experiments\cite{1.3,1.4} and has thereby been referred to as a piece of strong evidence for the medium being a deconfined plasma of quarks and gluons.\cite{1.5,1.6,1.7,1.8}  By the time of Quark Matter 2006 the frontier topic has moved beyond the suppression of single-particle distribution at high $p_T$ and into the correlation of hadrons on both the near side and the away side of jets.\cite{1.9,1.10}   The data on dihadron and trihadron correlations are currently analyzed for low and intermediate $p_T$, so the characteristic of hydrodynamical flow is involved in its interplay with semihard partons propagating through the medium.  The physics issues are therefore broadened from the medium effects on jets to include also the effect of jets on medium.  Theoretical studies of those problems can no longer be restricted to perturbative QCD that is reliable only at high $p_T$ or to hydrodynamics that is relevant only at low $p_T$.  In the absence of any theory based on first principles that is suitable for intermediate $p_T$, phenomenological modeling is thus inevitable.  A sample of some of the papers published before 2008 are given in Refs. [{\refcite{1.11}-\refcite{1.29}}].

Hadron correlation at intermediate $p_T$ involves essentially every complication that can be listed in heavy-ion collisions.  First, there is the characteristic of the medium created.  Then there is the hard or semihard scattering that generates partons propagating through it.  The interaction of those partons with the medium not only results in the degradation of the parton momentum, but also gives rise to ridges in association with trigger jets and to broad structure on the away side.  Those features observed are in the correlations among hadrons, so hadronization is an unavoidable subprocess that stands between the partonic subprocess and the detected hadrons.  Any realistic model must deal with all aspects of the various subprocesses involved.  Without an accurate description of hadronization, observed data on hadron correlation cannot be reliably related to the partonic origin of ridges and jet structure.

It is generally accepted that fragmentation is the hadronization subprocess at high $p_T$, as in lepton-initiated processes.  At intermediate and lower $p_T$ recombination or coalescence subprocess (ReCo) in heavy-ion collisions has been found to be more relevant.\cite{1.30,1.31,1.32,1.33}  Despite differences in detail, the three formulations of ReCo are physically very similar.  In Refs. [\refcite{1.31,1.32}] the descriptions are 3-dimensional and treat recombination and fragmentation as independent additive components of hadronization.  In Ref.\ [\refcite{1.33}] the formulation is 1D on the basis that acollinear partons have low probability of coalescence, and is simple enough to incorporate fragmentation as a component of recombination (of shower partons) so that there is a smooth transition from low to high $p_T$.  Since the discussions on jet-medium interaction in the main part of this review are based largely  on the formalism developed in Ref.\ [\refcite{1.33}] that emphasizes the role of shower partons at intermediate $p_T$, the background of the subject of recombination along that line is first summarized along with an outline of how non-trivial recombination functions are determined.  Some questions raised by critics, concerning such topics as entropy and how partons are turned into constituent quarks, are addressed.  More importantly, how shower partons are determined is discussed.

Large baryon-to-meson ratio observed in heavy-ion collisions is a signature of ReCo, since the physical reason for it to be higher than in fragmentation is the same in all three formulations.\cite{1.30,1.31,1.32}  The discussion here that follows the formulation of the recombination model (RM) by Hwa-Yang should not be taken to imply less significance of the other two, but only the limits of the scope of this review.  Considerable space is given to the topics of the Cronin effect (to correct a prevailing misconception) and to forward production at low and intermediate $p_T$ in Sec.\ 3.  The large B/M ratio observed at forward production cements the validity of recombination  so that one can move on to the main topic of jet-medium interaction.

The two aspects of the jet-medium interaction, namely, the effect of jets on the medium and that of the medium on jets, are discussed in Secs. 4 and 6, respectively.  In between those two sections we insert a section on azimuthal anisotropy because semihard jets can affect what is conventionally referred to as elliptic flow at low $p_T$ and also because ridge formation can depend on the trigger azimuth at intermediate $p_T$.  Much theoretical attention has been given in the past year to the phenomena of ridges on the near side and double hump on the away side of triggers at intermediate $p_T$.\cite{1.34}-\cite{1.48}  Our aim here is not to review the various approaches of those studies, but to give an overview of what has been accomplished on these topics in the RM.  The focus is necessary in order to cover a range of problems that depend on a reliable description of the hadronization subprocess. This review is complementary to the one given recently by Majumder,\cite{1.49} which emphasizes the region of $p_T$ much higher than what is considered here, so that factorized fragmentation can be applied.

Due to space limitations this review cannot go into the mathematical details of either the basic formalism or the specific problems.  Adequate referencing is provided to guide the interested reader to the original papers where details can be found.  The discussions will mainly be qualitative, thus rendering an opportunity to describe the motivations, assumptions and physical ideas that underlie the model calculations.  For example, the shower parton is an important ingredient in this approach that interpolates between what are soft (thermal-thermal) and hard (shower-shower), but we have neither space nor inclination to revisit the precise scheme in which the shower-parton distributions are derived from the fragmentation functions.  The concept of thermal-shower recombination and its application to intermediate-$p_T$ physics are more important than the numerical details.  Similarly, we emphasize the role that the ridges play (without triggers) in the inclusive distributions of single particles because of the pervasiveness of semihard scattering, the discussion of which can only be phenomenological.

Attempts are made to distinguish our approach from conceptions and interpretations that are generally regarded as conventional wisdom.  Some examples of what is conventional are:  (1) Cronin effect is due to initial-state transverse broadening; (2) large B/M ratio is anomalous; (3) azimuthal anisotropy is due to asymmetric high pressure gradient at early time; (4) recombination implies quark number scaling (QNS) of $v_2$; (5) dijets probe the medium interior.  In each case evidences are given to support an alternative interpretation.  In (4), it is the other way around: QNS confirms recombination but the breaking of QNS does not imply the failure of recombination.  Other topics are more current, so no standard views have been developed yet.  Indeed, there exist a wide variety of approaches to jet-medium interaction, and what is described here is only one among many possibilities.

\section{Hadronization by Recombination}

\subsection{A historical perspective}

In the 70s when inclusive cross sections were beginning to be measured in hadronic processes the only theoretical scheme to treat hadronization was fragmentation for lepton-initiated processes for which the interaction of quark were known to be the basic subprocess responsible for multiparticle production.  The same fragmentation process was applied also to the production of high-$p_T$ particles in hadronic collisions.\cite{2.1}  Local parton-hadron duality was also invoked as a way to avoid focusing on the issue of hadronization.\cite{2.2}  In dual parton model where color strings are stretched between quarks and diquarks, the fragmentation functions (FFs) are attached to the ends of the strings to materialize the partons to hadrons,  even if one of the ends is a diquark.\cite{2.3}  However, at low $p_T$ in $pp$ collisions quarks are not isolated objects in the parton model since there are gluons and wee-partons at small $x$,\cite{2.4} so the justification for the confinement of color flux to a narrow string is less cogent than at high $p_T$.   A more physically realistic description of hadronization seemed wanting.

The first serious alternative to fragmentation against the prevalent scheme for hadronization was the suggestion that pion production at low $p_T$ in $pp$ collisions can be treated by recombination.\cite{2.5}  The simple equation that describes it is
\begin{eqnarray}
 x {dN^{\pi}  \over  dx} = \int {dx_1  \over
x_1}{dx_2 \over  x_2} F_{q\bar{q}} (x_1, x_2) R^{\pi}(x_1, x_2, x) \ ,
\label{2.1}
\end{eqnarray}
where $F_{q\bar{q}} (x_1, x_2)$ is the $q\bar{q}$ distribution, taken to be the product $F_q(x_1)F_{\bar{q}}(x_2)$ of the $q$ and $\bar{q}$ distributions already known at the time among the parton distributions of a proton.  The recombination function (RF) $R^{\pi}(x_1, x_2, x) $ contains the momentum conserving $\delta(x_1 + x_2 - x)$ with a multiplicative factor that is constrained by the counting rule developed for quarks in hadrons.  That simple treatment of hadronization turned out to produce results that agreed with the existing data very well.

The next important step in solidifying the treatment of recombination is the detailed study of the RF.  If RF is circumscribed by the characteristics of the wave function of the hadron formed, then it should be related to the time-reversed process of describing the structure of that hadron.  In dealing with that relationship it also becomes clear that the distinction between partons and constituent quarks must be recognized and then bridged --- a problem that has puzzled some users of the RM even in recent years.  Since hadron structure is the basis for RF, it became essential to have a description of the constituents of a hadron in a way that interpolates between the hadronic scale and the partonic scale.  It is in the context of filling that need that the concept of valons was proposed.\cite{2.6}

The origin of the notion of constituent quarks (CQs) is rooted in solving the bound-state problem  of hadrons.  However, in describing the structure of a nucleon in deep inelastic scattering the role of CQs seems to be totally absent in the structure functions ${\cal F}$, such as $\nu W_2(x, Q^2)$.  The two descriptions are not merely due to the difference in reference frames,  CQs being in the rest frame, the partons in a high-momentum frame.  Also important is that the bound state is a problem at the hadronic scale, i.e.\ low $Q^2$, while deep inelastic scattering is at high $Q^2$.  The two aspects of the problem can be connected by the introduction of valons as the dressed valence quarks, i.e., each being a valence quark with its cloud of gluons and sea quarks which can be resolved only by high-$Q^2$ probes.  At low $Q^2$ the internal structure of a valon cannot be resolved, so a valon becomes what a CQ would be in the momentum-fraction variable in an infinite-momentum frame.  Thus the valon distribution in a hadron is the wave-function squared of the CQs, whose structure functions are described by pQCD at high $Q^2$.  Note that the usual description of $Q^2$-evolution by DGLAP has no prescription within the theory for the boundary condition at low $Q^2$.  That distribution at low $Q^2$ is precisely what the valon distribution specifies.  In summary, the structure function ${\cal F}^h(x,Q^2)$ of a hadron is a convolution of the valon distribution $G_{v/h}(y)$ and the structure function ${\cal F}^v(z,Q^2)$ of a valon\cite{2.7}
\begin{eqnarray}
 {\cal F}^h (x,Q^2)= \sum_{\nu} \int^{1}_{x} dy G_{v/h}(y)  {\cal F}^{v}\left({x/y} , Q^2 \right) \quad ,
\label{2.2}
\end{eqnarray}
where $y$ is the momentum fraction (not rapidity) of a valon in the hadron $h$.  The first description of the properties of $G_{v/h}(y)$ is given in [\refcite{2.6,2.7}], derived from the early data ${\cal F}^h (x,Q^2)$.  More recent determination of $G_{v/h}(y)$ is described in Ref.\ [\refcite{2.8}] where more modern parton distribution functions have been used.\cite{2.15}

$G_{v/h}(y)$ is the single-valon inclusive distribution in hadron $h$, and is the appropriate integral of the exclusive distribution, $G_{v/\pi}(y_1,y_2)$ for pion and
$G_{v/p}(y_1,y_2,y_3)$ for proton.  More specifically, $G_{v/\pi}(y_1,y_2)$ is the absolute square of the pion wave function $\left< v _1  (y_1) v _2 (y_2)|\pi\right>$ in the infinite-momentum frame.  Once we have that, it is trivial to get the RF for pion (i.e., by complex conjugation), since it is the time-reversed process.  Thus for pion and proton, we have
\begin{eqnarray}
R^{\pi}(x_1, x_2, x) = {x_1 x_2  \over x^2}\,G_{v /\pi}\left({x_1 \over x}, {x_2  \over x}  \right) \ ,
\label{2.3}
\end{eqnarray}
\begin{eqnarray}
R^p (x_1, x_2, x_3, x) = {x_1 x_2 x_3\over x^3}  G_{v/p} \left({x_1 \over
x}, {x_2 \over x},  {x_3 \over x}\right) \ ,
\label{2.4}
\end{eqnarray}
where the factors on the RHS are due to the fact that the RFs are invariant distributions defined in the phase space element $\Pi_i dx_i/x_i$, whereas  $G_{v/h}$ are non-invariant defined in $\Pi_i dy_i$, as seen in (\ref{2.2}).  The exclusive distribution $G_{v/h}$ contains the momentum conserving $\delta (\sum_i y_i - 1)$.  For pion there is nothing else, but for other hadrons the prefactors are given in Refs.\ [\refcite{2.8,2.16}].

Having determined the RF, the natural question next is how partons turn into valons before recombination in a scattering process.   Let us suppose that we can calculate the multi-parton distribution $F(x_1, x_2)$ for a $q$ and $\bar{q}$ moving in the same direction, whether at low or high $p_T$.  If their momentum vectors are not parallel, with relative transverse momentum larger than the inverse hadronic size, then the probability of recombination is negligible.  Relative longitudinal momentum need not be small, since the RF allows for the variation in the momentum fractions, just as the partons in a hadron can have various momentum fractions.  Now, as the $q$ and  $\bar{q}$ move out of the interaction region, they may undergo color mutation by soft gluon radiation as well as dress themselves with gluon emission and reabsorption with the possibility of creating virtual $q\bar{q}$ pairs, none of which can be made precise without a high $Q^2$ probe.  The net effect is that given enough time before hadronization the quarks convert themselves to valons with essentially the same original momenta, assuming that the energy loss in vacuum due to soft gluon radiation is negligible (even though color mutation is not negligible).  For that reason we may simply write $F (x_1, x_2)R (x_1, x_2, x)$ as multiplicative factors, as done in (\ref{2.1}), while treating $F (x_1, x_2)$ as the distribution of partons and $R (x_1, x_2, x)$ as the RF of valons.  The detail of this is explained in Ref.\ [\refcite{2.7}].

The question of entropy conservation has been raised at times, especially by those with experience in nuclear physics.  In elementary processes, such as $q + \bar{q} \to \pi$, unlike a nuclear process $p + n \to d$, the color degree of freedom is important.  Since a pion is colorless, the $q$ and $\bar{q}$  that recombine must have opposite color.  If they do not, they cannot travel  in vacuum without dragging a color flux tube behind them.  The most energy-efficient way for them to evolve is to emit soft gluons thereby mutating their color charges until the $q\bar{q}$  pair becomes colorless and recombine.  Such soft processes leave behind color degrees of freedom from the $q\bar{q}$  system whose entropy is consequently not conserved.  It is therefore pointless to pursue the question of entropy conservation in recombination, since the problem is uncalculable and puts no constraint on the kinematics of the formation of hadrons.  Besides, the entropy principle should not be applied locally.  A global consideration must recognize that the bulk volume is increasing during the hadronization process, and thus this compensates any decrease of local entropy density.

After the extensive discussion given above on the RF, we have come to the point of being able to assert that the main issue about recombination is the determination of the multi-parton distribution, such as $F_{q\bar{q}} (x_1, x_2)$ in (\ref{2.1}), of the quarks that recombine.  Related to that is the question about the role of gluons which have to hadronize also.  By moving the focus to the distributions of partons that hadronize, the investigation can then concentrate on the more relevant issues in heavy-ion collisions concerning the effect of the nuclear medium.

\subsection{Shower Partons}

At low $p_T$ in the forward direction the partons that recombine are closely related to the low-$Q^2$ partons in the projectile.  It is a subject to be discussed in a following section.  At intermediate and high $p_T$ the partons are divided into two types:  thermal (T) and shower (S).  The former contains the medium effect; the latter is due to semihard and hard scattered partons.  The consideration of shower partons is a unique feature of our approach to recombination, which is empowered by the possibility to include fragmentation process as SS or SSS recombination.  The jet-medium interaction is taken into account at the hadronization stage by TS recombination, although at an earlier stage the energy loss of the partons before emerging from the medium is another effect of the interaction  that is, of course, also important.  A quantitative theoretical study of that energy loss in realistic heavy-ion collisions at fixed centrality cannot be carried out and compared with data without a reliable description of hadronization.  At intermediate $p_T$ there is no evidence that fragmentation is applicable because the baryon/meson ratio would be too small, as we shall describe in the next section.

The fragmentation function (FF), $D(x)$, is a phenomenological quantity whose $Q^2$ evolution is calculable in pQCD; however, at some low $Q^2$ before evolution the distribution in $x$ is parametrized by fitting the data.  With that reality in mind it is reasonable to consider an alternative way of treating the FF, one that builds in more dynamical content by regarding fragmentation as  a recombination process.  That is, if we replace the LHS of Eq.  (\ref{2.1}) by the invariant function $x D^{\pi}_i (x)$, then the corresponding two-parton distribution in the integrand on the RHS is the product distributions of two shower partons in a jet initiated by a parton of type $i$.  To be specific, consider, for example, the fragmentation of gluon to pion
\begin{eqnarray}
 x D^{\pi}_g (x)  =  \int {dx_1 \over x_1} {dx_2 \over  x_2} S^{q}_g (x_1) S^{\bar{q}}_g\left({x_2 \over 1 - x_1}\right) R^{\pi}_{q\bar{q}}(x_1, x_2, x) \ ,
\label{2.5}
\end{eqnarray}
where the $q$ and $\bar{q}$ distributions in a shower initiated by the gluon are the same, but their momentum-fraction dependencies are such that if one $(x_1)$ is a leading quark, the other $(x_2)$ has to be from the remainder $(1 - x_1)$ of the parton pool.  With $ x D^{\pi}_g (x)$ being a phenomenological input, it is possible to solve (\ref{2.5}) numerically to obtain $S^{q}_g (z)$.  It has been shown in Ref.\ [\refcite{2.9}] that there are enough FFs known from analyzing leptonic processes to render feasible the determination of various shower parton distributions (SPDs), which are denoted collectively by $S^j_i$ with $i = q, \bar{q}, g$ and $j = q, s, \bar{q}, \bar{s}$, where $q$ can be either $u$ or $d$.  If in $i$ the initiating hard parton is an $s$ quark, it is treated as $q$.  That is not the case if $s$ is in the produced shower.  The parameterization of $S^j_i$ has the form
\begin{eqnarray}
S^j_i (z)  =  Az^a (1-z)^b (1 + cz^d) \ ,
\label{2.6}
\end{eqnarray}
where the dependence of $A$, $a$, etc., on $i$ and $j$ are given in a Table in Ref.\ [\refcite{2.9}].  Those parameters were determined from fitting the FFs at $Q = 10$ GeV, and have been used for all hadronization processes without further consideration of their dependence on $p_T$.  It should be recognized that those shower partons are not to be identified with the ones due to gluon radiation at very high virtuality calculable in pQCD, which is not applicable for the description of hadronization at low virtuality.

To sum up, in the conventional approach the FF is treated as a black box with a parton going in and a hadron going out, whereas in the RM we open up the black box and treat the outgoing hadron as the product of recombination of shower partons, whose distributions are to be determined from the FFs.  Once the SPDs are known, one can then consider the possibility that a shower parton may recombine with a thermal parton in the vicinity of a jet and thus give a more complete description of hadronization at intermediate-$p_T$ region, especially in the case of nuclear collisions.

The SPDs parametrized by (\ref{2.6}), being derived from the meson FFs, open up the question of what happens if, instead of $j\bar{j}' \to M$, three quarks in the shower recombine, e.g., $uud \to p$, or $uds \to \Lambda$.   A self-consistent scheme of hadronization would have to demand that the formation of baryons is a possibility in a fragmentation process and that the SPDs already determined should give an unambiguous prediction of what baryon FFs are.  The calculation has been carried out in Ref.\ [\refcite{2.10}] where the results for $g\to p$ and $g\to \Lambda$ in gluon jets are in good agreement with data\cite{2.11} without the use of any adjustable parameters.  To be able to relate meson and baryon FFs is an attribute of our formalism for hadronization that has not been achieved in other theoretical approaches, and provides further evidence that the SPDs are reliable for use at the hadronization scale.

\subsection{Parton distributions before recombination}

In the study of shower partons in a jet we have assumed the validity of the approximation that the fragmentation process is essentially one dimensional.  One may question whether the recombination process in a nuclear collision for a hadron produced at high $p_T$ may necessitate a 3D consideration, since two different length scales seem to be involved, one being that of the hadron produced, the other being the size of nuclear medium at hadronization time.  Indeed, recombination schemes formulated in 3D have been proposed, and various groups have independently found satisfactory results that are similar to one another.\cite{1.30,1.31,1.32}

The essence of recombination is, however, not in the 2D transverse plane normal to the direction of hadron momentum because if the coalescing parton momenta are not roughly parallel, then the relative momentum would have a large component in that transverse plane.  If that component is larger than the inverse of the hadron size, then the two (or three) partons cannot recombine.  Thus partons from regions of the nuclear medium that are far apart cannot form a hadron, rendering the concern over different length scales in the problem inessential.  Only collinear partons emanating from the same region of the dense medium can recombine.  For that reason the 1D formulation of recombination is adequate, as simple as expressed in (\ref{2.1}).  If one asks why the relative momentum can be large in the hadron direction, but not transverse to it, the answer lies in the foundation of the parton model where the momentum fraction can vary from 0 to 1, while the transverse parton momentum $k_T$ is limited to $\sim$ (hadron radius)$^{-1}$.    The RFs in (\ref{2.3}) and (\ref{2.4}) are related to the 1D wave function in that framework.

Having justified the 1D formulation of recombination, let us now focus on the distributions of the recombining partons at low $p_T$, and later at high $p_T$.  Since pQCD cannot be applied to multiparticle production at low $p_T$, our consideration of the problem is based on Feynman's parton model, which was originally proposed for hadron production at low $p_T$.\cite{2.4}  In a $pp$ collision there are valence and sea quarks and gluons whose $x$-distributions at low $Q^2$ are known.\cite{2.15}  Without hard scattering their momenta carry them forward, and they must hadronize in the fragmentation region of the initial proton.  RM has provided a quantitative treatment of the single-hadron inclusive distribution in $x_F$ not only for $pp$, but also for all realistic hadronic collisions.\cite{2.5,2.7,2.16}  What is to be remarked here is how gluons hadronize.  In an incident proton as in other hadrons, the gluons carry about half the momentum of the host.  Since gluons cannot hadronize by themselves, but can virtually turn to $q\bar{q}$ pairs in the sea, we require that all gluons be converted to the sea quarks (thus saturating the sea) before recombination.  This idea was originally suggested by Duke and Taylor,\cite{2.17} and was implemented quantitatively in the valon model in Refs.\ [\refcite{2.7,2.16}].  The simplest way to achieve that is to increase the normalization of the $q\bar{q}$ sea quarks without changing their $x$ distributions so that the total momentum of the valence (unchanged) and sea quarks (enhanced) exhausts the initial momentum of the hadron without any left over for gluons.  With the $F_{q\bar{q}} (x_1, x_2)$ thus obtained, the use of (\ref{2.1}) results in an inclusive $\pi$ distribution that agrees with data in both normalization and $x$ spectrum.\cite{2.7,2.16}  Using the appropriate valon distributions of pion and kaon, the success extends beyond $p \to \pi^{\pm}$ to $\pi^+ \to \pi^-$, $K^+ \to \pi^{\pm}$, and $\pi ^+ \to K^{\pm}$ in hadronic collisions at low $p_T$.  In nuclear collisions there is the additional complication arising from momentum degradation when partons traverse nuclear medium.  It is a subject that will be brought up in Sec.\ 3.4.

When $p_T$ is not small, then there has to be a semihard or hard scattering at the partonic level so that a parton with $k_T > 3$ GeV/c has to be created.  In that case shower partons are developed in addition to the thermal partons, so the partons before recombination can be separated into the following types:  TT+TS+SS for mesons and TTT+TTS+TSS+SSS for baryons.\cite{1.33}  The thermal partons have $k_T$ mainly $< 2$ GeV/c.  If one has a reliable scheme to calculate the thermal partons, then their $k_T$ distributions can, of course, be used in the recombination equation.  It does not mean that hydrodynamics is a necessary input in the RM.  In $pA$ collisions, for instance, hydrodynamics is not reliable, yet the Cronin effect can be understood in the RM for both proton and pion production without associating the effect with initial-state scattering --- a departure from the conventional thinking that will be discussed in the next section.  In most applications reviewed here, the distributions of thermal partons are determined from fitting the data at low $p_T$, and are then used in the RM to describe the behavior of hadrons at $p_T \ ^>_\sim\  3$ GeV/c.  When we consider correlation at a later section, careful attention will be given to the enhancement of thermal partons due to the energy loss of a semihard or hard parton passing through the nuclear medium.  It is only in the framework of a reliable hadronization scheme can one learn from the detected hadrons the nature of jet-medium interaction, as aspired in jet tomography.

\section{Large Baryon/Meson Ratios}
\subsection{Intermediate $p_T$ in heavy-ion collisions}

A well-known signature of the RM is that the baryon/meson (B/M) ratio is large --- larger than what is customarily expected in fragmentation.  The $p/\pi$ ratio of the FFs, i.e., $D_{p/q}(x)/D_{\pi/q}(x)$, is at most 0.2 at $x \simeq 0.3$, and is much lower at other values of $x$.\cite{2.11}  However, for inclusive distributions in heavy-ion collisions at RHIC the ratio $R_{p/\pi}$ is as large as $\sim 1$ at $p_T \simeq 3$ GeV/c, \cite{3.1,3.1a} as shown in Fig.\ 1.  Thus hadronization at intermediate $p_T$ cannot be due to parton fragmentation.  Three groups (TAM, Duke and Oregon) have studied the problem in the Recombination/Coalescence (ReCo) model\cite{1.30,1.31,1.32,1.33} and found large $R_{p/\pi}$ in agreement with the data.  The underlying reason that is common in all versions of ReCo is that for $p$ and $\pi$ at the same $p_T$ the three quarks that form the $p$ has average momentum $p_T/3$, while the
$q$ and $\bar{q}$ that form the $\pi$ has $p_T/2$.  Since parton distributions are suppressed severely at increasing $k_T$, there are more quarks at $p_T/3$ than at $p_T/2$, so the formation of proton is not at a disadvantage compared to that of a pion despite the difference in the RFs.  For either hadron the recombination process is at an advantage over fragmentation because of the addivity of momenta.  Fragmentation suffers from two penalties:  first, the initiating parton must have a momentum higher than
$p_T$, and second, the FFs are suppressed at any momentum fraction, more for proton than for pion.  Thus the yield from parton fragmentation is lower compared to that from parton recombination at intermediate $p_T$, even apart from the issue of B/M ratio.  When faced with the question why baryon production is so efficient, the proponents of pion fragmentation regard it as an anomaly.  Despite efforts to explain the enhancement in terms of baryon junction,\cite{3.2,3.3} the program has not been successful in establishing it as a viable mechanism for the formation of baryons.\cite{3.4}  From the point of view of ReCo there is nothing anomalous.


\begin{figure}[th]
\vspace{-1cm}
\centerline{\psfig{file=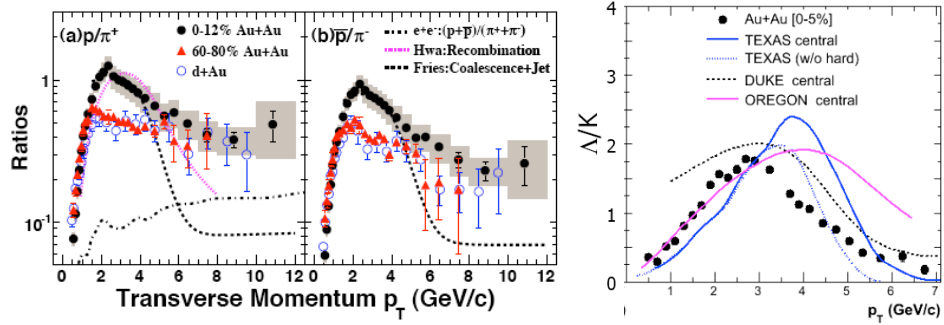,width=14cm}}
\vspace*{-5.7cm}
\caption{Comparison of baryon/meson ratio from STAR data to theoretical curves from ReCo.\cite{1.31,1.32,1.33} Left panel is from Ref.\ [\refcite{3.1}]; right panel is from Ref.\ [\refcite{3.1a}].}
\end{figure}

A simple way to understand the $p_T$ dependence of $R_{p/\pi}(p_T)$ is to consider the 1D formulation of ReCo given in Ref.\ [\refcite{1.33}], where the invariant distributions of meson and baryon production are expressed as
\begin{eqnarray}
p^0 {dN^M \over dp} = \int \left( \prod ^2_{i=1} {dq_i \over q_i} \right)F_{q\bar{q}} (q_1, q_2) R^M(q_1,q_2, p)
\label{3.1}
\end{eqnarray}
\begin{eqnarray}
p^0 {dN^B \over dp} = \int \left( \prod ^3_{i=1} {dq_i \over q_i} \right)F_{3q} (q_1, q_2, q_3) R^B(q_1,q_2, q_3, p)
\label{3.2}
\end{eqnarray}
in which all quarks are collinear with the hadron momentum $p$.  We assume that the rapidity $y$ is $\approx 0$, so the transverse momenta are the only essential variables, for which the subscripts $T$ of all momenta are therefore omitted, for brevity.  Mass effect at low $p_T$ renders the approximation poor and the 1D description inadequate.  However, in order to gain a transparent picture analytically, let us ignore those complications and assume provisionally that all hadrons are massless.  Then the experimental observation of exponential behavior of the $p_T$ distribution of pions at low $p_T$, i.e., $dN^{\pi}/pdp \propto \exp (-p/T)$, implies that the thermal partons behave as
\begin{eqnarray}
{\cal T} (q) = q {dN^{\rm th}\over dq} = Cqe^{-q/T} \quad ,
\label{3.3}
\end{eqnarray}
where $C$ has dimension (GeV)$^{-1}$, and $R^{\pi}$ given in (\ref{2.3}) is dimensionless.  When thermal partons dominate $F_{q\bar{q}}$ and $F_{3q}$, the multiparton distributions can be written as products:  ${\cal T}(q_1){\cal T}(q_2)$ and ${\cal T}(q_1){\cal T}(q_2){\cal T}(q_3)$, respectively.  It is then clear from the dimensionlessness of the quantities in (\ref{3.1}) and (\ref{3.2}) that with the proton distribution having $C^3$ dependence, as opposed to the pion distribution being $\propto C^2$, the $p/\pi$ ratio has the property
\begin{eqnarray}
R_{p/\pi} (p) = {dN^p/pdp \over dN^{\pi}/pdp}   \propto Cp  \quad ,
\label{3.4}
\end{eqnarray}
so long as thermal recombination dominates.  This linear rise with $p$ is the behavior seen in Fig.\ 1, although the mass effect of proton makes it less trivial in $p_T$.  Nevertheless, this simple feature is embodied in the more detailed computation until shower partons become important for $p_T > 3$ GeV/c.\cite{1.33}

From the above analysis which should apply to any baryon and meson, it follows that the ratios $\Lambda/K$ and $\Omega/\phi$ should also increase with $p_T$ in a way similar to $p/\pi$.  Such behaviors have indeed been observed by STAR,\cite{3.5,3.6} as have been obtained in theoretical calculation.\cite{3.7}  Taken altogether, it means that without TS and TTS recombination the B/M ratios would continue to rise with $p_T$.  But the data all show that the ratios peak at around $p_T \approx 3$ GeV/c.  In the RM the bend-over is due to the increase of the TS component of the meson earlier than the TTS component of the baryon,  since two thermal partons in the latter have more weight than the single thermal parton in the former.  The shower parton distribution ${\cal S}(q)$ in heavy-ion collisions is a convolution of the hard parton distribution $f_i(k)$ and the $S$ distribution derived from FF, discussed in Sec. 2.2, i.e.,
\begin{eqnarray}
{\cal S} (q) = \zeta  \sum_i \int dk kf_i(k) S_i (q/k) .
\label{3.4a}
\end{eqnarray}
$f_i(k)$ is the transverse-momentum distribution of hard parton $i$ at midrapidity and contains the shadowing effect of the parton distribution in nuclear collisions.  A simple parametrization of it is given in Ref.\  [\refcite{3.7a}] as follows
\begin{eqnarray}
f_i (k) = K {A \over (1 + k/B)^n}  \quad ,
\label{3.4b}
\end{eqnarray}
where $K = 2.5$ and $A, B, n$ are tabulated for each parton type $i$ for nuclear collisions at RHIC and LHC.  The parameter $\zeta = 0.07$ is the average suppression factor that can be related to the nuclear modification factor $R_{AA}$, and was denoted by $\xi$ in Ref.\ [\refcite{1.33}] and other references thereafter.  Since $f_i(k)$ has a power-law dependence on $k$, so does ${\cal S}(q_2)$ on $q_2$ in contrast to the exponential behavior of the thermal partons, ${\cal T}(q_1)$.  This upward bending of ${\cal S}(q_2)$ relative to ${\cal T}(q_1)$ is the beginning of the dominance of TS and TTS components over TT and TTT components, resulting in a peak in the B/M ratio at around $p_T \sim 3$ GeV/c.  Detailed descriptions of these calculations are given in Ref.\ [\refcite{1.33,3.7}].

We add here that the effort made to consider the shower partons before recombination is motivated by our concern that a hard parton with high virtuality cannot hadronize by coalescing with a soft parton with low virtuality.  The introduction of shower partons is our way to bring the effects of hard scattering to the hadronization scale.  At the same time the formalism does not exclude fragmentation by a hard parton, since SS and SSS recombination at high $p_T$ are equivalent to fragmentation but in a language that has dynamical content at the hadronization scale.

One could ask how  the RM can be applied reliably in the intermediate-$p_T$ region before the shower partons were introduced.  The approach adopted in Ref.\  [\refcite{1.30}] does not involve the determination of the hard parton distribution by perturbative calculation, but uses the pion data as input to extract the parton distribution at the hadronization scale at all $p_T$ in the framework of the RM.  It is on the basis of the extracted parton distribution (which must in hindsight contain the shower partons) that the proton inclusive distribution is calculated.  Thus the procedure is  self-consistent.  The result is that the $p/\pi$ ratio is large at $p_T \sim 3$ GeV/c in agreement with data; furthermore, it was a prediction that the ratio would decrease as $p_T$ increases beyond 3 GeV/c, as confirmed later by data shown in Fig.\ 1.

\subsection{Cronin effect}

The conventional explanation of the Cronin effect,\cite{3.8} i.e., the enhancement of hadron spectra at intermediate $p_T$ in $pA$ collisions with increasing nuclear size, is that it is due to multiple scattering of projectile partons as they propagate through the target nucleus, thus acquiring transverse momenta, and that a moderately large-$k_T$ parton hadronizes by fragmentation.\cite{3.9}  The emphasis has been on the transverse broadening of the parton in the initial-state interaction (ISI) and not on the final-state interaction (FSI).  In fact, the Cronin effect has become synonymous to ISI effect in certain circles.  However, that line of interpretation ignores another part of the original discovery\cite{3.8} where the $A$ dependence of hadrons produced in $pA$ collisions, when parameterized as
\begin{eqnarray}
{dN \over dp_T} \left(pA \to hX\right)  \propto A^{\alpha _h (p_T)}  \quad ,
\label{3.5}
\end{eqnarray}
has the property that $\alpha _p (p_T)>\alpha _{\pi} (p_T)$ for all $p_T$ measured.  That experimental result alone is sufficient to invalidate the application of fragmentation to the hadronization process, since if the $A$ dependence in (\ref{3.5}) arises mainly from the ISI, where the multiply-scattered parton picks up it $k_T$, then the transverse broadening of that parton should have no knowledge of whether the parton would hadronize into a proton or a pion, so $\alpha _h$ should be independent of the hadron type $h$.

A modern version of the Cronin effect is given in terms of the central-to-peripheral nuclear modification factor for $dAu$ collisions at midrapidity
\begin{eqnarray}
R^h_{CP} \left(p_T \right) = {\left(1/N^C_{coll} \right)dN^h/ p_T  dp_T (C) \over \left(1/N^P_{coll} \right)dN^h/ p_T  dp_T (P)}   \quad ,
\label{3.6}
\end{eqnarray}
where $C$ and $P$ denote central and peripheral, respectively, and $N_{coll}$ is the average number of inelastic $NN$ collisions.  If hadronization is by fragmentation, which is a factorizable subprocess, the FFs for any given $h$ should cancel in the ratio of (\ref{3.6}), so $R^h_{CP}$ should be independent of $h$.  However, the data show that $R^p_{CP} (p_T)> R^{\pi}_{CP} (p_T)$ for all $p_T > 1$ GeV/c when $C =$ 0-20 \% and $P =$ 60-90 \% centralities.\cite{3.10}  See Fig.\ 2.  Clearly ISI is not able to explain this phenomenon, which strongly suggests the medium-dependence of hadronization.   The data further indicate that the  $p_T$ dependence of $R^h_{CP} \left(p_T \right)$ peaks at $p_T \sim 3$ GeV/c for both $p$ and $\pi$, reminiscent of the $p/\pi$ ratio at fixed centrality in $AuAu$ collisions although the $C/P$ ratio for $dAu$ collisions is distinctly different.

Hadron production at intermediate $p_T$ and $\eta \sim 0$ in $dAu$ collisions can be treated in the RM in a similar way as for $AuAu$ collisions.  Although no hot and dense medium is produced in a $dAu$ collision, so thermal partons are not generated in the same sense as in $AuAu$ collisions, nevertheless soft partons are present to give rise to the low-$p_T$ hadrons.  For notational uniformity we continue to refer to them as thermal partons.  We apply the same formalism developed in Ref.\ [\refcite{1.33}] to the $dAu$ problem and consider the TT+TS+SS contributions to $\pi$ production (TTT+TTS+TSS+SSS for $p$).  The thermal ${\cal T}$ distribution is determined by fitting the $p_T$ spectra at $p_T < 1$ GeV/c for each centrality; the shower-parton ${\cal S}$ distribution is calculated as before but without nuclear suppression.  Unlike the dense thermal system created in $AuAu$ collisions, the ${\cal T}$ distribution in this case is weaker;  its parameters $C$ and $T$ (inverse slope) that correspond to the ones in Eq.\ (\ref{3.3}) are smaller.  Furthermore, $C$ decreases with increasing peripherality, while $T$ remains unchanged at 0.21 GeV/c.\cite{3.11}  Thus thermal-shower recombination becomes important at $p_T \ ^>_\sim\ 1$ GeV/c, which is earlier than in $AuAu$ collisions.  As a consequence, $R^{\pi}_{CP} (p_T) $ becomes $> 1$ at $p_T > 1$ GeV/c.  That is the Cronin effect, but not due to ISI.  The same situation occurs for proton production, only stronger.\cite{3.12}  The calculated results for the inclusive distributions of both $\pi$ and $p$ agree well with data at all centralities, hence also $R^{\pi}_{CP} (p_T) $ and $R^{p}_{CP} (p_T) $.  Fig.\ 2 shows $R^{h}_{CP} (p_T) $ for $C =$ 0-20 \% and $P =$ 60-90 \%  in $dAu$ collisions for $h = \pi$ and $p$;\cite{3.10}  the lines are the results obtained in the RM.\cite{3.11,3.12}  The reason for  $R^p_{CP}> R^{\pi}_{CP}$ can again be traced to 3-quark recombination for $p$ and only 2 quarks for $\pi$.  When $p_T$ is large, fragmentation dominates (i.e. SS and SSS), and both $R^{h}_{CP}$ approach 1, since FFs cancel and the yields are normalized by $N_{coll}$.  No exotic mechanism need be invoked to explain the $p$ production process.  FSI alone is sufficient to provide the underlying physics for the Cronin effect.

\begin{figure}[th]
\vspace{-1cm}
\centerline{\psfig{file=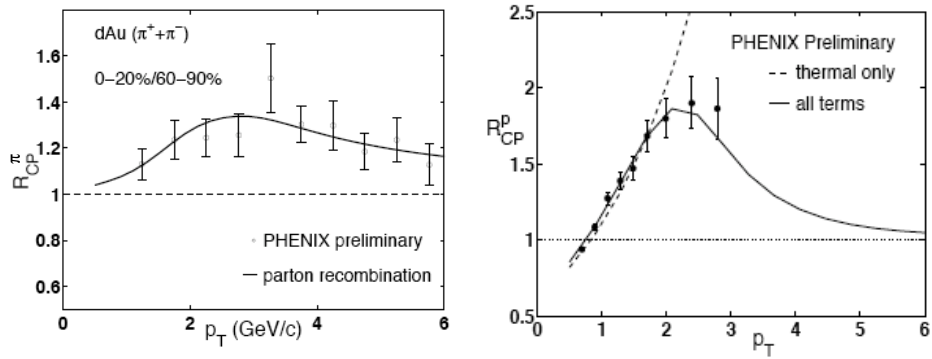,width=14cm}}
\vspace*{-5.3cm}
\caption{Central-to-peripheral ratios for the production of pion (left panel) and proton (right panel) in $dAu$ collisions. Data are from Ref.\ [\refcite{3.10}] and lines are from Refs.\ [\refcite{3.11,3.12}].}
\end{figure}

\subsection{Forward production in $dAu$ collisions}

Hadron production at forward rapidities in $dAu$ collisions was regarded as a fertile ground for exposing the physics of ISI, especially saturation physics,\cite{3.13,3.14} since the nuclear effect in the deuteron fragmentation region was thought to cause minimal FSI.  It was further thought that the difference in nuclear media for the $Au$ side $(\eta < 0)$ and the $d$ side ($\eta > 0$) would lead to backward-forward asymmetry in particle yield in such a way as to reveal a transition in basic physics from multiple scattering in ISI for $\eta\ ^<_\sim\ 0$  to gluon saturation for $\eta > 0$.  The observation by BRAHMS\cite{3.15} that $R_{CP}$ decreases with increasing $\eta$ was regarded as an indication supporting that view.\cite{3.16}  That line of thinking, however, assumes that FSI is invariant under changes in $\eta$ so that any dependence on $\eta$ observed is a direct signal from ISI.  Such an assumption is inconsistent with the result of a study of forward production in $dAu$ collisions in the RM, where both $R_{CP} (p_T, \eta) $ and $ R_{B/F} ({p_T})$ are shown to be well reproduced by considering FSI only.\cite{3.17}  Any inference on ISI from the data must first perform a subtraction of the effect of FSI, and just as in the case of the Cronin effect there is essentially nothing left after the subtraction.

In Sec. 3.2 the Cronin effect at midrapidity $(\eta \approx 0)$ is considered.  The extension to $\eta > 0$ along the same line involves no new physics.  However, it is necessary to determine the $\eta$ dependencies of the soft and hard parton spectra at various centralities.  For the soft partons, use is made of the data on $dN/d\eta$ to modify the normalization of ${\cal T}(q,\eta)$ already determined at $\eta = 0$.  For the hard partons, modified parametrizations of their distributions $f_i(k_T,\eta)$ are obtained from leading order minijet calculations using the CTEQ5 pdf \cite{3.18} and the EKS98 shadowing.\cite{3.19}  A notable feature of the result is that $f_i(k_T,\eta)$ falls rapidly with $k_T$ as $\eta$ increases, especially near the kinematical boundary $k_T = 8.13$ GeV/c and $\eta = 3.2$.  Thus TS and SS components are negligible compared to TT at large $\eta$ for any $p_T$ and any centrality, even though the TT component is exponentially suppressed.  In central collisions there is the additional suppression due to momentum degraduation of the forward partons going through the nuclear medium of the target $Au$.  Putting the various features together leads to the ratio $R_{CP} (p_T, \eta) $ shown in Fig.\ 3(a), where the data are from Ref.\ [\refcite{3.15}] and the curves from the calculation in Ref.\ [\refcite{3.17}].  It is evident that the decrease of $R_{CP} (p_T, \eta) $ at $p_T > 2$ GeV/c as $\eta$ increased from $\eta = 0$ to $\eta = 3.2$ is well reproduced in the RM.  Only one new parameter is introduced to describe the centrality and $\eta$ dependence of the inverse slope $T$ of the soft partons, but no new physics has been added.  The suppression of $R_{CP} (p_T, \eta) $ at $\eta > 1$ is due mainly to the reduction of the density of soft partons in the forward direction, where hard partons are suppressed.

Extending the consideration to the backward region and using the same $T(\eta)$ extrapolated to $\eta < 0$, the backward/forward ratio of the yield can be calculated.\cite{3.17}  For $\eta = \pm 0.75$ corresponding to the data of STAR at $0.5 < |\eta| < 1.0 $ and 0-20\% centrality,\cite{3.20} the calculated result on $R_{B/F}$ for $\pi^+ + \pi^- + p + \bar{p}$ is shown by the solid line in Fig.\ 3(b).  While it agrees with the data very well for $p_T < 2$ GeV/c, it is noticably lower than the data for all charged particles for $p_T > 2$ GeV/c.  However, more recent data on $R_{B/F} (p_T)$ for $\pi^++\pi^-+p+\bar{p}$,\cite{3.21} shown in the inset of  Fig.\ 3(b), exhibit excellent agreement with the same theoretical curve that should be regarded as a prediction.

\begin{figure}[th]
\vspace{-1cm}
\centerline{\psfig{file=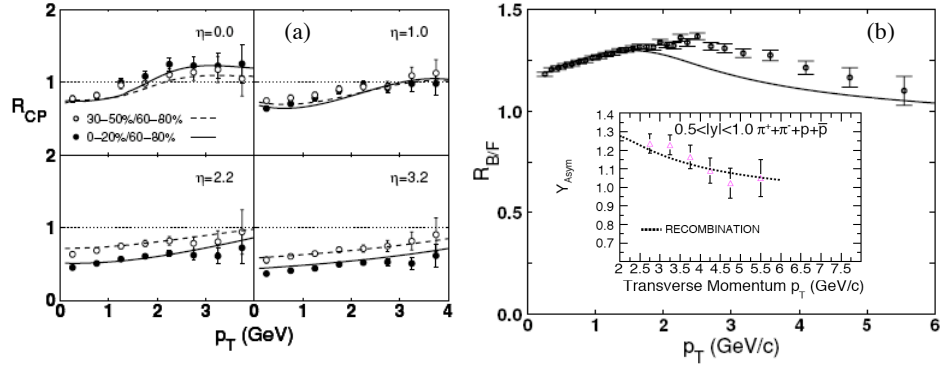,width=14cm}}
\vspace*{-5.3cm}
\caption{(a) $R_{CP}(p_T,\eta)$ for forward production. Data are from Ref.\ [\refcite{3.15}] and lines from Ref.\ [\refcite{3.17}]. (b) Back/Forward ratio $R_{B/F}(p_T)$ for $0.5<|\eta|<1.0$ that shows agreement of theoretical result \cite{3.17} with preliminary data \cite{3.20} for $p_T<2$ GeV/c but not for higher $p_T$, but later data \cite{3.21} (inset) show agreement for $p_T>2$ GeV/c also.}
\end{figure}

The fact that $R_{B/F}$ is $> 1$ for all $p_T$ measured may be regarded as a proof against initial transverse broadening of partons, since forward partons of $d$ have more nuclear matter of $Au$ to go through than the backward partons of $Au$.  Thus if ISI is responsible for the acquisition of $p_T$ of the final-state hadrons, then  $R_{B/F}$ should be $< 1$.   The data clearly indicate otherwise.

\subsection{Forward production in $AuAu$ collisions}

Theoretical study of hadron production in the forward direction in heavy-ion collision is a difficult problem for several reasons.  The parton momentum distribution at low $Q^2$ and large momentum fraction $x$ in nuclear collisions is hard to determine, especially when momentum degradation that accounts for what is called ``baryon stopping'' cannot be ignored.  Furthermore, degradation of high-momentum partons in the nuclear medium implies the regeneration of soft partons at lower $x$; that is hard to treat also.  The use of data as input to constrain unknown parameters is unavoidable; however, existent data have their own limitations.  Measurement at fixed $\eta$ cannot be used to provide information on $x_F$ dependence unless $p_T$ is known.\cite{3.22}  Measurement of both $\eta$ and $p_T$ has been limited to charged hadrons\cite{3.23} that cannot easily be separated into baryons and mesons.  For these various reasons forward production in $AA$ collisions has not been an active area of theoretical investigation.  However, there are gross features at large $\eta$ that suggest important physics at play and deserve explanation.

PHOBOS data show that particles are detected at $\eta ' > 0$ where $\eta '$ is the shifted pseudorapidity defined by $\eta ' = \eta - y_{\rm beam}$.\cite{3.22}  It is significant because it suggests that if $\langle p_T \rangle$ is not too small, it  corresponds to $x_F > 1$, where $x_F  = (p_T/m_p) e^{\eta '}$.  Instead of violation of momentum conservation, the interpretations in the RM is that a proton can be produced in the $x_F > 1$ region, if three quarks from three different nucleons in the projectile nucleus, each with $x_i < 1$, recombine to form a nucleon with $x_F  = \sum_i x_i > 1$.\cite{3.24}  That kinematical region is referred to as transfragmentation region (TFR), which is not accessible, if hadronization is by fragmentation.  The theoretical calculation in the RM involves an unknown parameter, $\kappa$, which quantifies the degree of momentum degradation of low-$k_T$ partons, in the forward direction.  For $\kappa$ in a reasonable range, not only can nucleons be produced continuously across the $x_F$ boundary, but also can $p/\pi$ ratio attain an amazingly large value.\cite{3.24}

BRAHMS has determined the $p_T$ distribution of all charged particles at $\eta = 3.2$.\cite{3.23}  For $\langle p_T \rangle = 1$ GeV/c, the corresponding values of $x_F$ for pion and proton are, respectively, 0.4 and 0.54.  Taking the preliminary value of the $\bar{p}/p$ ratio at 0.05 into account, it is possible to estimate the value of $\kappa$ and then calculate the $p_T$ distribution of $p + \bar{p} + \pi^+ + \pi^- $.\cite{3.26}  The $p/\pi$ ratio was predicted to be $\sim 1$ at $p_T \sim 1$ GeV/c.  However, at QM2008 the more recent data on $R_{\bar{p}/p}$ was reported to have  a lower value at 0.02 \cite{3.27} and on $R_{p/\pi}$ a higher value at $\sim 4$ at $p_T \sim 1$ GeV/c.\cite{3.28}  Those new data prompted a reexamination of the problem in the RM; with appropriate changes in the treatment of degradation, regeneration and transverse momentum, the very large $p/\pi$ ratio can be understood.\cite{3.29}

Since $p$, $\bar{p}$ and $ \pi$ production at large $\eta$ depends sensitively on $q$ and $\bar{q}$ distributions, which in turn depend strongly on the dynamical process of momentum degradation and soft-parton regeneration (the parameterization of which requires phenomenological inputs), the procedure in Ref.\ [\refcite{3.29}] is to use $R_{p/\pi}$ and $R_{\bar{p}/p}$ as input in order to determine $\kappa$ and then calculate the $x$ distributions of the hadrons.  At fixed $\eta$ the $x$ and $p_T$ distributions are related.  It turns out that the result on the $x$ distribution leads to a large contribution to the $p_T$ distribution of $R_{p/\pi} (p_T)$ shown by the dashed line in Fig.\ 4(a).   The additional enhancement shown by the solid line arises from the mass dependence of the inverse slopes $T_h$ due to flow.  While the ratio  $R_{p/\pi}$ is insensitive to the absolute normalizations of the yields, the inclusive distribution of all charged particles is not.  In Fig.\ 4(b) is shown the good agreement between the calculated result and the data in both normalization and shape with no extra parameters beyond $\kappa$ already fixed.

\begin{figure}[th]
\vspace{-1cm}
\centerline{\psfig{file=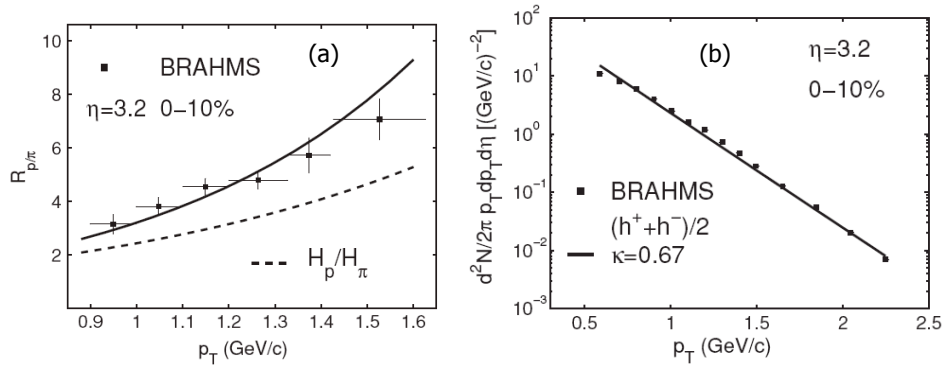,width=14cm}}
\vspace*{-5.3cm}
\caption{(a) Proton/pion ratio in forward production at $\eta=3.2$ showing agreement between data\cite{3.28} and solid line from the RM;\cite{3.29} the dashed line is the contribution from the longitudinal components at fixed $\eta$. (b) Comparison of the $p_T$ distribution of charged particles at $\eta=3.2$ from BRAHMS\cite{3.23} with calculated result from the RM.\cite{3.29}}
\end{figure}

It should be noted that the $p/\pi$ ratio, shown in Fig.\ 4(a), is extremely large at $\eta = 3.2$ and modest $p_T < 2$ GeV/c.  The underlying physics is clearly the suppression of $\bar{q}$ at medium $x$ and the enhancement of $p$ due to $3q$ recombination, where the (valence) quarks are from three different nucleons in the projectile.  No other hadronization mechanisms are known to be able to reproduce the data on the large $R_{p/\pi}$ at large $\eta$.

\subsection{Recombination of adjacent jets at LHC}

So far we have considered only the physics at RHIC energies and the recombination of thermal and shower partons, either between them or among themselves.  At RHIC high-$p_T$ jets are rare, so the shower partons are from one jet at most in an event.  At LHC, however, high-$p_T$ jets are copiously produced for $p_T < 20$ GeV/c.  When the jet density is high, the recombination of shower partons in neighboring jets becomes more probable and can make a significant contribution to the spectra of hadrons in the $10 < p_T < 20$ GeV/c range, high by RHIC standard, but intermediate at LHC.  If that turns out to be true, then a remarkable signature is predicted and is easily measurable:  the $p/\pi$ ratio will be huge, perhaps as high as 20.\cite{3.30}

If a hard parton of momentum $k_T$ is produced, shower partons in its jet with momenta $q_i$ are limited by the constraint $\sum_i q_i < k_T$, so that the recombination of those shower partons can produce a hadron with momentum $p_T$ not exceeding $k_T$.  However, if there are two adjacent jets with hard-parton momenta $k_{1T}$ and $k_{2T}$, then to form a hadron at $p_T$ from shower partons in those two jets, neither $k_{1T}$ nor $k_{2T}$ need to be larger than $p_T$, so the rate of such a process would be higher.  Furthermore, to form a proton at $p_T$ the shower parton $q_i$ can be lower than those for pion formation at the same $p_T$, so $k_{iT}$ can be even lower.  Thus $R_{p/\pi}$ in 2-jet recombination can be much higher than the ratio in 1-jet fragmentation.

The probability for 2-jet recombination, however, also depends on the overlap of jet cones, since the coalescing shower partons must be nearly collinear.  That overlap decreases with increasing $k_{iT}$, so there is a suppression factor in the SS or SSS recombination integral that depends on the widths of the jet cones.  Using some reasonable estimates on all the factors involved, it is found that
$R_{p/\pi}$ can be between 5 and 20 in the range $10 < p_T < 20$ GeV/c, decreasing with increasing $p_T$.\cite{3.30}  Although exact numbers are unreliable, the approximate value of $R_{p/\pi}$ is about 2 orders of magnitude higher than what is expected in the usual scenario of fragmentation from single hard partons.

The origin of the large $R_{p/\pi}$ at LHC discussed above is basically the same as that for forward production in $AuAu$ collisions at RHIC.  In both cases it is the multi-source supply of the recombining partons that enhances the proton production.  At large $p_T$ at LHC there are more than one jet going in the same direction; at large $p_L$ at RHIC there are more than one nucleon going in the forward direction.  In the latter case we already have data supporting our view that $R_{p/\pi}$ should be large as shown in Fig.\ 3(a).  It would be surprising that our prediction of large $R_{p/\pi}$ at LHC turns out to be untrue.

\section{Ridgeology -- Phenomenology of Ridges}

In the previous section the topics of discussion have been exclusively on the single-particle distributions in various regions of phase space.  Everywhere it is found that the B/M ratio is large when $p_T$ is in the intermediate range.  We now consider two-particle correlations, on which there is a wealth of data as a result of the general consensus in both the experimental and theoretical communities that more can be learned about the dense medium when one studies the system's effect on (and response to) penetrating probes.  The strong interaction between energetic partons and the medium they traverse, resulting in jet quenching, is the underlying physics that can be revealed in the jet tomography program.\cite{1.2,4.2}  To calibrate the medium effect theoretically, it is necessary to have a reliable framework in which to do calculation from first principles, and that is perturbative QCD.  Although many studies in pQCD have been carried out to learn about the modification of jets in dense medium in various approximation schemes,\cite{4.3,1.49} they are mainly concerned with the effect of the medium on jets at high $p_T$, and the results can only be compared with data on single-particle distributions, such as $R_{AA}(p_T)$.   The response of the medium to the passage of hard partons is not what can be calculated in pQCD, since it involves soft physics.  That is, however, the physical origin of most of the characteristics in the correlation data.  An understanding of that response is one of the objectives of studying correlations.  Without the reliable theory to describe correlation, especially at low to intermediate $p_T$ where abundant data exist, it becomes necessary to use phenomenological models to relate various features of correlation.  When all the features can consistently be explained in the framework of a model, then one may feel that a few parameters are a small price to pay for the elucidation of the jet-medium interaction.

On two-particle correlation the most active area in recent years has been the use of triggers at intermediate or high $p_T$ to select a restricted class of events and the observation of associated particles at various values of $\eta$ and $\phi$ relative to the trigger.\cite{1.01,4.4}  Among the new features found, the discovery of ridges on the near side has stimulated extensive interest and activities.\cite{4.6}  We review in this section only those aspects in which recombination plays an important role, which in turn makes inferences on the origin of the ridges.  We start with a summary of the experimental facts.

\subsection{Experimental features of ridges}

The distribution of particles associated with a trigger at intermediate $p_T$ exhibits a peak at small $\Delta\eta$ and $\Delta \phi$ sitting on top of a ridge that has a wide range in $\Delta \eta$, where $\Delta\eta$ and $\Delta \phi$ are, respectively, the differences of $\eta$ and $\phi$ of the associated particle from those of the trigger.\cite{4.4,4.7}  A 2D correlation function in ($\Delta\eta, \Delta \phi$) first shown by Putschke\cite{4.7} at QM06 is reproduced here in Fig.\ 5(a).  STAR has been able to separate the ridge $(R)$ from the peak $(J)$, where $J$ refers to Jet, although both are features associated with jets.  The structure shown in Fig.\ 5(a) is for $3 < p^{\rm trig}_T < 4$ GeV/c and $p^{\rm assoc}_T  > 2$ GeV/c in central $AuAu$ collisions.  The ridge yield integrated over $\Delta\eta$ and $\Delta \phi$ decreases with decreasing $N_{\rm part}$, until it vanishes at the lowest $N_{\rm part}$ corresponding to $pp$ collisions, so $R$ depends strongly on the nuclear medium.  That is not the case with $J$.  On the other hand, $R$ is also strongly correlated to jet production, since the ridge yield is insensitive to $p^{\rm trig}_T $.  Thus the ridge is a manifestation of jet-medium interaction.

Putschke further showed\cite{4.7} that the ridge yield is exponential in its dependence on $p^{\rm assoc}_T$ and that the slope in the semi-log plot is essentially independent of $p^{\rm trig}_T $.  That is shown by the solid lines in Fig.\ 5(b).  The inverse slope parameterized by $T$ is slightly higher than $T_0$ of the inclusive distribution, also shown in that figure.  Since the $p_T$ range in that figure is between 2 and 4 GeV/c, we know from single-particle distribution that the shape of the inclusive spectrum is at the transition from pure exponential on the low side to power-law behavior on the high side.   The last data point at  $ p^{\rm assoc}_T = 4$ GeV/c  being above the straight line is an indication of that.  Thus the value $T_0$ of the pure exponential part for the bulk is lower than what that
straight line suggests.  The exponential behavior of $R$ should be taken to mean that the particles in the ridge are emitted from a thermal source.  Usually thermal partons are regarded as begin uncorrelated.  In the case of $R$ they are all correlated to the semihard parton that initiates the jet.  We thus interpret the observed characteristics as indicating that the ridge is from a thermal source at $T$, enhanced by the energy lost by the semihard parton transversing the medium at $T_0$.

\begin{figure}[th]
\vspace{-1cm}
\centerline{\psfig{file=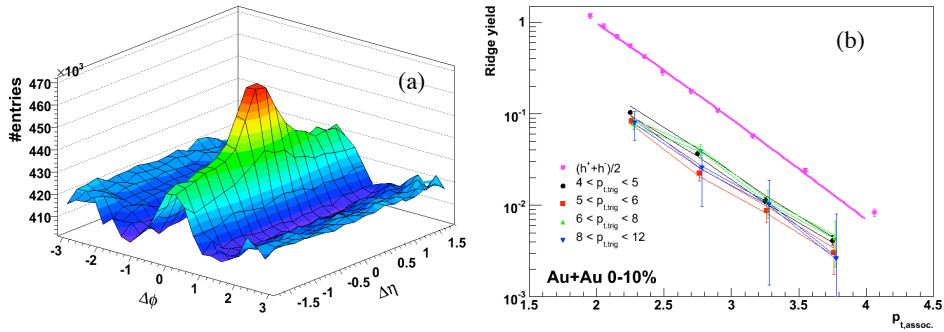,width=14.5cm}}
\vspace*{-5.8cm}
\caption{(a) Jet structure from Ref.\ [\refcite{4.7}] for charged particles associated with a trigger. (b) Dependence of ridge yield\cite{4.7} on $p_T^{\rm assoc}$ for various $p_T^{\rm trig}$.}
\end{figure}

The B/M ratio of particles in the ridge is found to be even higher than the same ratio of the inclusive distributions in $AuAu$ collisions at 200 GeV.  More specifically, $(p+\bar{p})/(\pi^+ + \pi^-)$ in $R$ for  $p^{\rm trig}_T  > 4$ GeV/c and $2 < p^{\rm assoc}_T < p^{\rm trig}_T$ is about 1 at $p_T = 4$ GeV/c.\cite{4.8}  In contrast, that ratio in $J$ is more than 5 times lower.  There is indication that the $\Lambda /K$ ratio in the ridge is just as large.\cite{3.5}  As discussed in Sec.\ 3, it is hard to find any way to explain the large B/M ratio outside the framework of recombination.  Since the exponential behavior in $p_T$ implies the hadronization of thermal partons, the application of recombination very naturally gives rise to large B/M ratio, as we have seen in Sec.\ 3.1.

Putting together all the experimental features discussed above, we can construct a coherent picture of the dynamical origin of the $R$ and $J$ components of the jet structure, although no part of it can be rigorously proved for lack of a calculationally effective theory of soft physics.  There are several stages of the dynamical process.

\begin{description}
\item[(a)]\quad A hard or semihard scattering takes place in the medium resulting in a parton directed outward in the transverse plane at midrapidity.  Because of energy loss to the medium, those originating in the interior are not able to transverse the medium as effectively as those created near the surface.  That leads to trigger bias.
\item[(b)]\quad Whatever the nature of the jet-medium interaction is, the energy lost from the semihard parton goes to the enhancement of the thermal energy of the partons in the near vicinity of the passing trajectory.  Those enhanced thermal partons are swept by the local collective movement outward whether or not the flow can be described by equilibrated hydrodynamics initially.
\item[(c)] \quad Since the initial scattering takes place at $|\eta| < 0.7$, which is the pseudorapidity range of the trigger acceptance, the shower (S) partons associated with the jet are restricted to the same range of $\eta$.  However, the enhanced thermal partons that interact strongly with the medium can be carried by the high-$\eta$ initial partons that they encounter on the way out and be boosted to higher $\eta$.  Thus the distribution of the enhanced thermal partons is elongated in $\Delta \eta$, but not in $\Delta \phi$ because the expansion of the bulk system is in longitudinal and radial directions, not in the azimuthal direction.  Consequently, the hadronization of the enhanced thermal partons has the shape of a ridge.
\item[(d)]\quad  In terms of recombination the ridge is formed by TT and TTT recombination, while the peak $J$ is formed largely by TS and TTS (or TSS) recombination, and possibly also by fragmentation (SS and SSS), depending on $p_T$ and centrality.  Since the $J$ component involves S, it is restricted to a narrow cone in $\Delta \eta$ and $\Delta \phi$.
\end{description}

An initial attempt to incorporate all these properties in the RM was made in Ref.\ [\refcite{1.13}] before the ridge data were reported in QM06.\cite{4.7}  By the time of QM08 ridgeology has become an intensely studied subject, as evidence by the talks in Ref.\ [\refcite{4.6}].

\subsection{Recombination of enchanced thermal partons}

Although the properties of ridges described in the above subsection are derived from events with triggers, it should be recognized that ridges are present with or without triggers.  That is because the ridges are induced by semihard scattering which can take place whether or not a hadron in a chosen $p_T$ range is used to select a subset of events.  Experimentally, it is known that the peak and ridge structure is seen in auto-correlation where no triggers are used.\cite{4.11}  The implication is that the ridge hadrons are pervasive and are always present in the single-particle spectra.

Hard scattering of partons can occur at all virtuality $Q^2$, with increasing probability at lower and lower $Q^2$.  When the parton $k_T$ is $< 3$ GeV/c, the rate of such semihard scattering can be high, while the time scale involved is low enough ($\sim 0.1$ fm/c) to be sensitive to the initial spatial configuration of the collision system.  Thus for noncentral collisions there can be nontrivial $\phi$ dependence, which we shall discuss in Sec.\ 5.  Hadron formation does not take place until much later, so it is important to bear in mind the two time scales involved in ridgeology.  Ridges are the hadronization products of enhanced thermal partons at late time, which are stimulated by semihard parton created at early time.  In the absence of a theoretical framework to calculate the degree of enhancement due to energy loss, we extract the characteristics of the thermal distributions from the data.  Although hydrodynamics may be a valid description of the collective flow after local thermal equilibrium is established, it does not take semihard scattering into consideration and assumes fast thermaliztion without firmly grounded justification.  If the semihard scattering occurs in the interior of the dense medium, the energy of the scattered parton is dissipated in the medium and contributes to the thermalization of the bulk ($B$).  That process may take some time to complete.  If the semihard scattering occurs near the surface of the medium, its effect can be detected as $J + R$ in these events selected by a trigger with the trigger direction not far from the local flow direction, a point to be discussed in more detail later in Sec.\ 4.4.  Inclusive distribution averages over all events without triggers, including all manifestation of hard and semihard scatterings; hence, it is the sum of $B+R+J$.  Since $J$ is associated with the shower partons $(S)$, we identify $J$ with the recombination of TS+SS for the mesons and TTS+TSS+SSS for the baryons, leaving TT+TTT  for $B+R$.  Thus the exponential behavior of the thermal partons is revealed in the exponential behavior of $B+R$ in $p_T$, for which we emphasize the inclusion of the ridge contribution to the inclusive distribution.

In noncentral collisions the ridges are not produced uniformly throughout all azimuth,\cite{4.12} so $dN/dp_T$ that averages over all $\phi$ has varying proportions of $B$ and $R$ contributions depending on centrality.  To be certain that we can get a measure of the $R$ contribution independent of $\phi$, we focus on only the most central collisions in this and the next subsections.  Continuing to use the notation $k_T$ for the transverse momentum of the semihard parton at the point of creation in the medium, $q_T$ for that at the point of exit from the medium, and $p_T$ for the hadron outside, we have for thermal partons the distribution given in (\ref{3.3}) just before recombination.  Our first point to stress here is that the inverse slope $T$ in (\ref{3.3}) includes the effects of both $B$ and $R$.  Putting that expression into (\ref{3.1}) where one takes 
\begin{eqnarray}
F_{q\bar{q}}(q_{1_T}, q_{2_T}) = {\cal T}(q_{1_T}){\cal T}(q_{2_T})
\label{4.1}
\end{eqnarray}
and being more explicit with the RF for pion in (\ref{2.3}), i.e.,
\begin{eqnarray}
R^{\pi}(q_{1_T}, q_{2_T}, p_T) = {q_{1_T}q_{2_T} \over p^2_T} \delta \left({q_{1_T} \over p_T} + {q_{2_T} \over p_T} - 1 \right) \ ,
\label{4.2}
\end{eqnarray}
one obtains\cite{1.33}
\begin{eqnarray}
{dN^{B + R}_{\pi}\over p_Tdp_T} = {C^2 \over 6} e^{-p_T/T} \ ,
\label{4.3}
\end{eqnarray}
although in 2004 no one was aware of the existence of ridges.  From the data\cite{4.13} for identified hadrons, one can fit the $\pi ^+$ distribution for 0-5\% centrality in the range $1 < p_T < 3$ GeV/c and get $T = 0.3$ \rm GeV/c.  This value is slightly lower than the one given in Ref.\ [\refcite{4.7}] which takes the slope of the inclusive distribution in the range $2 < p_T < 3.5$ GeV/c.  Ref.\ [\refcite{4.13}] provides data for $K$ and $p$ also, which have the same value of $T$ as above for $1 < p_T < 3$ GeV/c, thus confirming that the exponential behaviors of the hadronic spectra can be traced to the common thermal distribution in (\ref{3.3}) through recombination.  At lower $p_T$ the spectra for $K$ and $p$ deviate from exponential behavior because of mass effect, which can largely be taken into account by using $E_T$ instead of $p_T$, where
\begin{eqnarray}
E_T(p_T) = m_T - m_0, \quad m_T = \left(p^2_T +m_0^2\right)^{1/2} \quad ,
\label{4.4}
\end{eqnarray}
$m_0$ being the hadron rest mass.  Thus we write for all hadrons
\begin{eqnarray}
{dN^{B+R}_h \over p_Tdp_T} = A_h(p_T)e^{-E_T(p_T)/T'_h} ,
\label{4.5}
\end{eqnarray}
where $A_{\pi}(p_T) = C^2/6$ is a constant for pion, but for proton $A_p(p_T) = C^3A_0 p^2_T/p_0$ where $A_0$ is a numerical factor that arises from the wave functions (valon distribution) of the proton.\cite{1.33}  Note that the inverse slope is now denoted by $T'_h$, since the data\cite{4.13} show dependence on hadron type when the distributions are plotted as functions of $E_T$.  Furthermore, $T'_p$  is found to depend on centrality, which is a feature that can be understood in the RM as being due to the non-factorizability of the thermal parton distributions of $uud$ at very peripheral collisions where the density of thermal partons is low.\cite{4.14}  For central collisions, $T'_p = 0.35$ GeV.  We summarize the empirical results for $\pi$ and $p$ as follows:
\begin{eqnarray}
T'_{\pi} = 0.3 \, {\rm GeV},
\label{4.6}
\end{eqnarray}
\begin{eqnarray}
T'_p = 0.35 (1 - 0.5 c) \,{\rm GeV},
\label{4.7}
\end{eqnarray}
where $c$ denotes \% centrality, e.\ g., $c = 0.1$ for 10 \%.  We shall hereafter use $T'_h$ to denote the inverse slope in $E_T$ for $B + R$, and $T_h$ for that in $E_T$ for $B$ only, i.e.,
\begin{eqnarray}
{dN^{B}_h \over  p_Tdp_T}= B_h(p_T) = A_h(p_T)e^{-E_T(p_T)/T_h} .
\label{4.8}
\end{eqnarray}

It is hard to find data that describes the bulk contribution only, since the effect of semihard scattering cannot easily be filtered out.  Indeed, as $p_T \to 0$, there is no operational way without using trigger to distinguish $B(p_T)$ from all inclusive.  For that reason the prefactor $A_h(p_T)$ in (\ref{4.8}) is the same as that in (\ref{4.5}).  In events with trigger above a threshold momentum, semihard partons with lower momenta than that threshold can contribute to $R$; it becomes a part of the background, which is experimentally treated as $B$.  Thus the only meaningful way to isolate $R$ quantitatively is by use of correlation, while accepting the difficulty of separating $B$ and $R$ outside the momentum ranges where the correlated particles are measured.    Another way of stating that attitude is to accept the experimental paradigm of regarding the mixed events as a measure of the background (hence, by definition, the bulk), and treating $R$ as only that associated with a trigger.  Our cautionary point to make is that such a background can contain untriggered ridges.  In practice, one can take the difference between (\ref{4.5}) and (\ref{4.8}) and identify it as the ridge yield

\begin{eqnarray}
{dN^{R}_h \over p_Tdp_T}= R_h(p_T) = A_h(p_T)e^{-E_T(p_T)/T'_h} \left[1 - e ^{-E_T(p_T)/T''_h} \right],
\label{4.9}
\end{eqnarray}
where
\begin{eqnarray}
{1 \over T''_h} =  {1 \over T_h} - {1 \over T'_h} = {\Delta T_h \over T_h T'_h}, \qquad \Delta T_h = T'_h - T_h  .
\label{4.10}
\end{eqnarray}
If $ \Delta T_h  \ll T_h$, then the quantity in the square bracket makes a small correction to the $\exp \left[ -E_T(p_T)/T'_h\right]$ behavior, and one can determine $T'_h$ from the data.  The only data available that address the ridge distribution are in Ref.\ [\refcite{4.7}] where the associated particles are in the range $2 < p^{\rm assoc}_T < 4$ GeV/c, exhibiting an approximately exponential behavior.  It is shown in Ref.\ [\refcite{4.14}] by using the data for trigger momentum in the range $4 < p^{\rm trig}_T < 5$ GeV/c that with $\Delta T_h = 45$ MeV in (\ref{4.10}) the ridge distributions can be well fitted.  The expression for $R_h(p_T)$ in (\ref{4.9}) has no explicit dependence on $p^{\rm trig}_T$, as is roughly the case with the data.  It does have strong dependence on $p^{\rm assoc}_T$, which is $p_T$ in (\ref{4.9}).  Experimental exploration of the lower $p^{\rm assoc}_T$ region would provide further validation that (\ref{4.9}) needs.  The physics basis for that distribution is the recombination of thermal partons given in (\ref{3.3}).

\subsection{Trigger from the ridge}

We have discussed above the observation of ridge in triggered events, but to have a trigger from the ridge seems to put the horse behind the cart.  There must be a phenomenological motivation for that role reversal.

Let us start with the single-pion inclusive distribution that shows an exponential decrease in $p_T$ followed by a power-law behavior.  The boundary between the two regions is at $\sim 2$ GeV/c.  We have associated the exponential region to TT recombination and the power-law region to TS+SS.  We have also discussed the contribution to T from the enhanced thermal partons arising from the medium response to semihard partons.  In order to be able to investigate the TT component better without the interference from the shower contribution so that one can examine the $B +  R$ components cleanly, it would be desirable to be able to push the TS+SS components out of the way.  That is not possible with the light $u$ and $d$ quarks, but not impossible with the $s$ quark, since the heavier quark is suppressed in hard scattering.

If one observes the hadrons formed from only the $s$ quarks, either $\phi$ or $\Omega$, one finds exponential behavior at all $p_T$ measured, which in the case $\Omega$ extends to as high as $5.5$ GeV/c.\cite{4.15}-\cite{4.21}  The absence of any indication of up-bending of the distributions clearly suggests
that the source of the $s$ quarks is thermal in nature and that no shower partons participate in the formation of $\phi$ and  $\Omega$.  That problem is studied in Ref.\ [\refcite{4.22}] along with $K$ and $\Lambda$ production.  Indeed, the data can be well reproduced by TT for $\phi$, TTT for $\Omega$, TT+TS for $K$, and TTT+TTS+TSS for $\Lambda$.

Since $s$ quarks in S make insignificant contribution to $\Omega$ production for $p_T < 6$ GeV/c, and since thermal partons are uncorrelated, it is reasonable to expect that the $\Omega$ observed has no correlated particles.  It was therefore predicted\cite{4.23} that if $\Omega$ in $3 < p_T < 6$ GeV/c is treated as a trigger particle, there should be no associated particles appearing as a peak in $\Delta\phi$ on the near side.  Within a year that prediction was falsified by a report at QM06 showing that there is a near-side peak after background subtraction in $\Delta \phi$ for $\Omega$ in the range $2.5 < p^{\rm trig}_T < 4.5$ GeV/c in central $AuAu$ collisions with charged particles in the range $1.5 < p^{\rm assoc}_T <p^{\rm trig}_T$.\cite{4.24}  The data created a dilemma: is $\Omega$ created by a jet or not?  If it is, why is the $p_T$ distribution strictly exponential with no hint of jet characteristics?  If it is not, why is there a $\Delta \phi$ peak in azimuthal correlation?  The dilemma became known as the $\Omega$ puzzle.\cite{4.25}

The resolution of that puzzle is in the recognition that both the trigger $\Omega$ and the associated particles are in the ridge, first conjectured in Ref.\ [\refcite{4.25}] and later quantified in Ref.\  [\refcite{4.26}].  Jets are involved, since without jets there can be no ridge.  But not all jet structures exhibit a prominent peak above a ridge.  It depends on the trigger particle and the ranges of $p^{\rm trig}_T$ and $p^{\rm assoc}_T$.  Consider the jet yield compared to the ridge yield at $3 < p^{\rm trig}_T < 6$ GeV/c in $AuAu$ collisions at 0-10\% centrality.\cite{3.5,4.28}  The $J/R$ ratio at $p^{\rm assoc}_T \sim 1.2$ GeV/c decreases as the trigger particle goes from $h$ to $K^0_S$ and then to $\Lambda/\bar{\Lambda}$.  For $\Lambda/\bar{\Lambda}$ trigger and unidentified charged $h$ associated, the $J/R$ ratio is $\leq 10^{-1}$ for $|\Delta \eta |_J < 0.7$ and $|\Delta \eta |_R < 1.7$.  Since the $\Lambda/\bar{\Lambda}$ trigger must contain an $s$ quark which is absent in the shower, the participating $s$ quark must be a thermal parton. For $p^{\rm trig}_T$ near 3 GeV/c, thermal $s$ quark around $1$ GeV/c or less can be quite abundant, and the initiating semihard parton need not be very hard.  With $p^{\rm assoc}_T$ as low as $\sim 1$ GeV/c, the light hadrons in $R$ dominate over those in $J$, so $J/R$ is small.  As the strangeness in the trigger increases, more thermal $s$ quarks are involved with less dependence on shower.  $J/R$ is likely to be even smaller, although present data with $\Omega$ trigger lack statistics to show the $\Delta \eta$ distribution.  If the jet structure shows mainly a ridge with negligible peak $(J)$ in
$\Delta \eta$, we have referred to it as a phantom jet,\cite{4.25} i.e., a Jet-less jet.  The corresponding $\Delta \phi$  distribution should then be dominantly $R$.  Since the initiating semihard partons are either gluon or light quarks, the usual jet structure may still be seen if the trigger particle is ordinary.  But for $\Omega$ trigger the structure is very different.

Now, we can address the $\Omega$ puzzle.  The enhanced thermal partons generated by the semihard parton contain $s$ as well as $u$ and $d$ quarks.  Three $s$ quarks anywhere in the ridge can recombine to form a trigger $\Omega$.  Other quarks, in particular the light quarks, in the ridge can form associated particles.  The pool of enhanced thermal partons are all correlated to the semihard parton in every event selected by the $\Omega$ trigger, so the associated particles are all restricted to $|\Delta \phi|<1$.  Since the $s$ quarks that form the $\Omega$ are thermal, the $\Omega$ spectrum in $p_T$ is exponential.  Hence the $\Omega$ puzzle is solved.  The trigger can be from the ridge.

The above description outlines a detailed calculation of the $\Delta \phi$ distribution of charged hadrons produced in association with $\Omega$.\cite{4.26}  The background is calculated in the RM using previous parametrizations, and the height agrees with the data.  That is important, since the ridge signal is less than 4\% of the background height.  Two adjustable parameters are used to fit the $\Delta \phi$ distribution  of the ridge, but then the yield/trigger is calculated as a function of $p^{\rm trig}_T$ without further unknowns in the model, and the result is in agreement with the data,\cite{3.5,4.28} as shown in Fig.\ 6.  The solid and dashed curves in the two panels of that figure are the results of the calculation using two values of the strength of enhanced thermal partons that differ by only 1\%, yet the height of the ridge varies by about 20\%.  That is because the ridge is the difference between large numbers of $B + R$ and $B$.  Such accuracy is beyond the scope of any dynamical theory to achieve.  The phenomenological approach adopted in Ref.\ [\refcite{4.26}] has been the only one that offers a quantitative understanding of the $\Omega$ problem.

\begin{figure}[th]
\vspace{-1cm}
\centerline{\psfig{file=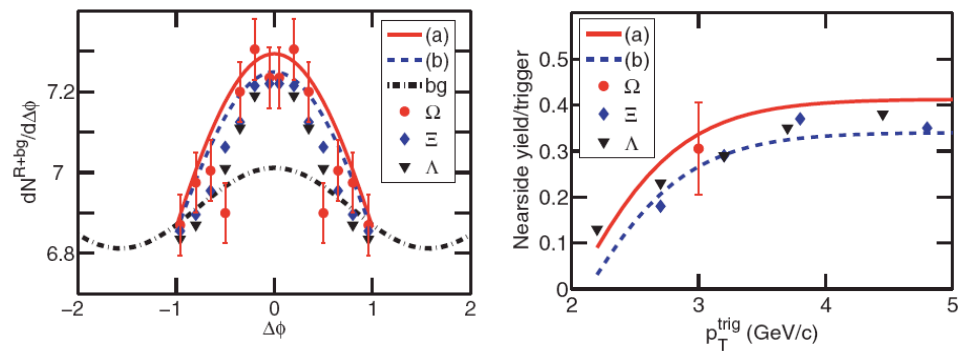,width=14.5cm}}
\vspace*{-5.8cm}
\caption{(a) Left panel: Data\cite{4.28} on associated particle distribution in $\Delta\phi$ for three hyperon triggers at $2.5<p_T ^{\rm trig}< 4.5$ GeV/c and  $1.5<p_T ^{\rm assoc}< p_T ^{\rm trig}$. (b) Right panel: Near-side ridge yield associated with $\Omega$ trigger. The lines are from calculations in the RM\cite{4.26} for particles associated with $\Omega$ trigger, with the solid and dashed lines for slightly different normalization constants of the Gaussian peak. }
\end{figure}

Recently, PHENIX has shown data\cite{4.29} that can be interpreted as support for the notion of trigger from the ridge.  At $p^{\rm trig}_T < 4$ GeV/c (for unidentified trigger) the per-trigger yield of the associated particles is found to be less than expected from fragmentation, and the ``dilution'' effect is attributed to the increase of the number of triggers due to soft processes.  In our language fragmentation is the SS component, and the TS component due to medium effect gives   an increase already over fragmentation for $p_T < 6$ GeV/c.  The additional dilution effect at $p_T < 4$ GeV/c is due to TT recombination, which is enhanced by the ridge contribution.

\subsection{Dependence of ridge formation on trigger azimuth}

So far our consideration in ridgeology has been concerned mainly with the dependence on $p_T$.  Now, let us turn to azimuthal correlation, although our discussion of the main topic of azimuthal anisotropy of single-particle distributions is deferred to the following section.  Our present focus is on the correlation between the directions of the trigger and ridge particles in the transverse plane, a topic that can be discussed prior to $\phi$ anistropy because it is mainly a problem in ridge formation.  The subject was stimulated by the report\cite{4.12} that the ridge yield depends on the azimuthal angle  $\phi _s$ between the trigger angle $\phi _T$ and the reaction plane $\Psi _{RP}$, even for nearly central collisions, but especially for noncentral collisions.

Since geometry is an important factor that influences the $\phi _s$ dependence, it is necessary to treat carefully the initial configuration of the problem:  (a)  the point of origin $(x_0, y_0)$ of the semihard parton in the almond-shaped overlap region, (b) the angle $\phi _s$ of the parton's trajectory, (c) the density of the medium, $D(x, y)$, along that trajectory, and (d) the point of exit from the medium.  In the approximation that the medium does not expand very much during the time that the semihard parton near the surface traverses the medium, it is not difficult to calculate the path length, but it is much more difficult to calculate the energy loss that depends on $D(x, y)$.  Even if there is a reliable way to account for the effect of the medium on the semihard parton, there is no known way to translate that to the effect of the parton on the medium.  The enhancement of the thermal partons that lead to the ridge particles takes time to develop, during which the medium expands.  Local flow direction depends on where the enhanced thermal partons are in the overlap, which evolves into elliptical geometry.  Each of the various parts of the process can be represented by a factor that can be expressed in terms of variables that have reasonable physical relevance.  Without entering into the details that are described in Ref.\ [\refcite{4.30}] we can outline the essence here.

Let $P(x_0, y_0, t)$ denote the probability of detecting a parton emerging from the medium, where $t$ is the path length measured from the initial point $(x_0, y_0)$ to the surface along a straight-line trajectory at angle $\phi _s$.  Let $C(\psi(x,y), \phi _s)$ be a function that describes the correlation between the enhanced thermal partons along the flow direction $\psi(x,y)$ and the semihard parton direction $\phi _s$.  Finally, let $\Gamma (x, y, \phi)$ describe the fluctuation of the angle $\phi$ of a ridge particle from the average flow direction.  For fixed $(x_0, y_0)$ and $\phi _s$, the ridge-particle distribution in $\phi$ is then
\begin{eqnarray}
R(\phi, \phi _s, x_0, y_0) = NP (x_0, y_0, t)t   \int^1_0 d \xi D (x_{\xi}, y_{\xi}) C (x_{\xi}, y_{\xi}, \phi _s) \Gamma (x_{\xi}, y_{\xi}, \phi) \ ,
\label{4.11}
\end{eqnarray}
where $N$ is a normalization constant related to the rate of production of the ridge particle and $\xi$ is the fraction of the path length $t$ along the trajectory starting at $0$ at $(x_0, y_0)$ and ending at $1$.  For the observed distribution it is necessary to integrate over all $(x_0, y_0)$.  Not every semihard parton included in that integration gets out of the medium to generate a particle that triggers the event.  The ridge distribution per trigger is therefore normalized by the probability of the ridge-generating parton emerging from the medium
\begin{eqnarray}
R(\phi, \phi _s)  = {\int dx_0dy_0 R(\phi, \phi _s, x_0, y_0)\over \int dx_0dy_0 P(x_0, y_0, t)} \ ,
\label{4.12}
\end{eqnarray}
where the integration is over the entire region of initial overlap.  There is no explicit dependence on
$p^{\rm trig}_T$ and $p^{\rm assoc}_T$ in (\ref{4.11}), but the parameters specifying the different factors in the equation do.  We leave the momenta fixed in the ranges $3 < p^{\rm trig}_T <4$ GeV/c and $1.5 < p^{\rm assoc}_T < 2.0$ GeV/c, as specified in the experiment,\cite{4.12} and focus only on the dependencies on $\phi$ and $\phi _s$.

The most important factor in (\ref{4.11}) is $C(x, y, \phi_s)$, which is parametrized as
\begin{eqnarray}
C(x, y, \phi_s) = \exp \left[- {\left(\phi_s - \psi (x,y)\right)^ 2\over 2 \lambda}\right]
\label{4.13}
\end{eqnarray}
where $\lambda$ specifies the Gaussian width of the correlation between the directions of semihard parton $\phi_s$ and local flow $ \psi (x,y)$.  If  $\psi (x,y)$ is close to $\phi _s$ for most of the points $(x, y)$ along the trajectory, then the thermal partons enhanced by successive soft emissions are carried by the flow along the same direction.  The effects reinforce one another and lead to the formation of a ridge in a narrow cone.  On the other hand, if $ \psi (x,y)$ is very different from $\phi _s$, then the enhanced soft partons are dispersed over a range of surface area, so their hadronization leads to no pronounced effect.  That is the essence of the correlated emission model (CEM).\cite{4.30}

The probability $P(x_0, y_0, t)$ depends, in addition to the nuclear thickness at $(x_0, y_0)$, the survival probability $S(t)$, which is exponentially behaved in $t$.  Most of the semihard partons that can emerge from the medium are created in a layer near the surface.  In Ref.\ [\refcite{4.30}] the thickness of that layer is set to be approximately $R_A/4$.  A more detailed study of that will be described in Sec. 6 below.  The fluctuation distribution $\Gamma (x, y, \phi)$ turns out not to influence the final result in any sensitive way.

A pictorial description of the origin of $\phi_s$ dependence of ridge yield in CEM is shown in the left panel of Fig.\ 7, where (a) $\phi_s \sim 0$ and (b) $\phi_s \sim 70^{\circ}$ in noncentral collisions.  The trigger directions $\phi_s$ are represented by the thin arrows, while the flow direction $\psi (x, y)$ are depicted by the thick arrows that are normal to the surface.  When the two types of arrows match, the reinforcement leads to a strong ridge.  In (a) the matching condition is met where the density is higher than in (b) where the matching arrows occur near the top of the overlap.  Thus the former case for $\phi_s \sim 0$ has higher ridge yield than in the latter case for larger $\phi _s$.  That difference becomes
more drastic as the collision centrality becomes more peripheral.  The results on the ridge yield integrated over $\phi$ are compared to the data in the right panel of Fig.\ 7.  In (a) of that panel the normalization is adjusted to fit the data point at the lowest $\phi _s$ and the shape of the $\phi _s$ dependence is adjusted by varying the parameter $\lambda$ in (\ref{4.13}).  The value determined is $\lambda = 0.11$; it corresponds to a correlation cone of half-width $\sim 20 ^{\circ}$.  The curve in part (b) of that panel for 20-60\% centrality is obtained without any further adjustment.  Although the agreement with data is not perfect, the curve does reproduce the trend of steep descend, followed by a region of flatness or even a small up-bending.  The behavior at large $\phi _s$  is due to the additional contribution of semihard partons from the left-half ellipse at high $y_0$.

\begin{figure}[th]
\vspace{-1cm}
\centerline{\psfig{file=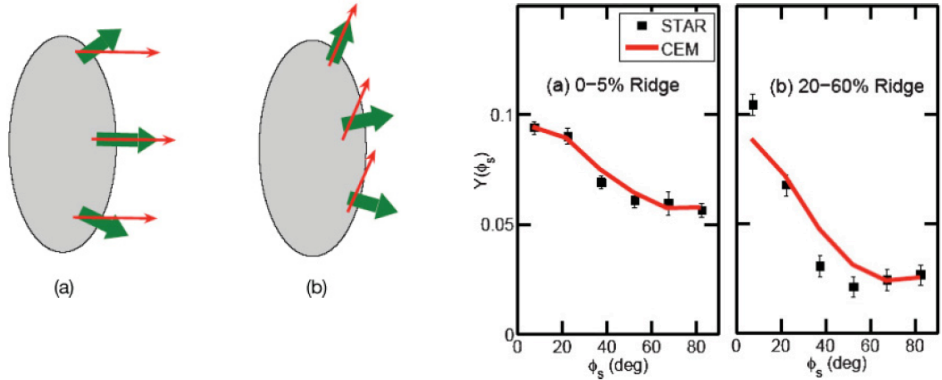,width=14.5cm}}
\vspace*{-5.6cm}
\caption{Left panel: schematic sketches of trigger directions (thin arrows) and flow directions (thick arrows) in noncentral collisions. Right panel:  Data\cite{4.12} on the ridge yield/trigger vs $\phi_s$ for (a) 0-5\% and (b) 20-60\% centrality. The lines are from CEM\cite{4.30} with common normalization adjusted to fit the left-most data point in (a). }
\end{figure}

Without being integrated over $\phi$, the ridge distribution $R(\phi, \phi_s)$ given by (\ref{4.11}) and (\ref{4.12}) describes the dependence of the ridge yield on $\Delta \phi = \phi - \phi_s$.  An interesting discovery upon careful study of  $R(\phi, \phi_s)$ is that its dependence on $\Delta \phi$ is not symmetric across the $\Delta \phi = 0$ point,  depending on the sign of $\phi_s$.  In Fig.\ 8(a)
 is shown its behavior for six positive values of $\phi _s$.  Not shown are the mirror images of those curves across $\Delta \phi = 0$ for corresponding negative values of $\phi _s$.  Since the measurement\cite{4.12} averages over both positive and negative values of $\phi _s$, there is no observable asymmetry in the data.  The reason for the asymmetry in Fig.\ 8(a) is that for $\phi _s > 0$ most of the points along the trajectory have $\psi (x, y) < \phi _s$, since $\psi (x, y)$ is generally normal to the surface.  Hence, the ridge particles are mostly at $\Delta \phi < 0$.  The reverse is true for $\phi _s < 0$.  The left shift of the peaks in $\Delta \phi$ in Fig.\ 8(a) has been confirmed by STAR recently.\cite{4.31}

 \begin{figure}[th]
\vspace{-1cm}
\centerline{\psfig{file=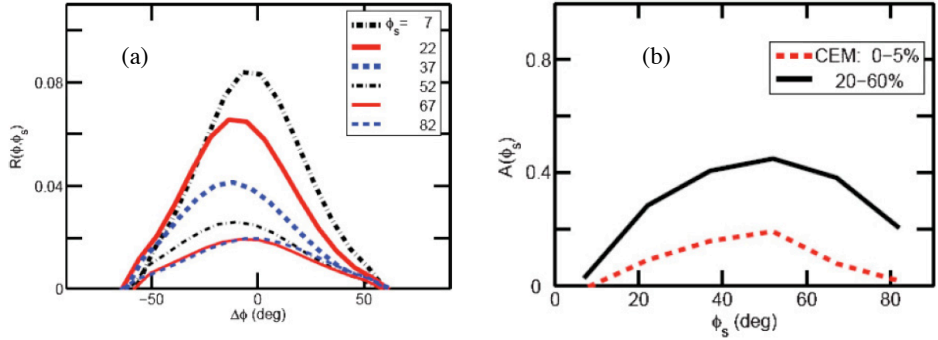,width=14.5cm}}
\vspace*{-6cm}
\caption{(a) Left shift of the ridge yield in $\Delta\phi$ for $\phi_s>0$ in CEM.\cite{4.30}  (b) Inside-outside asymmetry function defined in Eq.\ (\ref{4.15}).}
\end{figure}

Another way to test the asymmetry is to measure a quantity $A(\phi _s)$ called inside-outside asymmetry function.\cite{4.30}  To that end, define for $0 < \phi _s < \pi/2$
\begin{eqnarray}
Y_+ (\phi_s) = \int^{\phi_s }_ {\phi_s  - 1} d \phi R(\phi, \phi _s), \qquad Y_- (\phi_s) = \int^{\phi_s +1}_ {\phi_s} d \phi R(\phi, \phi _s)
\label{4.14}
\end{eqnarray}
and for $- \pi /2 < \phi _s < 0$ reverse the definition.  Then for any $\phi_s$ define
\begin{eqnarray}
A(\phi_s) = {Y_+ (\phi_s)  - Y_- (\phi_s)  \over Y_+ (\phi_s)  + Y_- (\phi_s)} \quad .
 \label{4.15}
\end{eqnarray}
This asymmetry function should always be positive whether there is a left shift for $\phi _s < 0$ or a right shift for $\phi _s > 0$.  By reflection symmetry it vanishes at $\phi _s = 0$ and $\pm \pi/2$.  In CEM the properties of $A(\phi_s)$ at two centralities are shown in Fig.\ 8(b).   STAR reported very recently at QM09 that such an asymmetry has indeed been found to exist in the data on the ridges for various $\phi_s$.\cite{4.31,4.32}

\section{Azimuthal Anisotropy}

The azimuthal dependence of single-particle distribution has been studied ever since the beginning of heavy-ion physics.\cite{5.1}-\cite{5.4}  Hydrodynamical model at low $p_T$\cite{5.5}-\cite{5.8a} and ReCo at intermediate $p_T$\cite{1.31,1.32,5.9,5.10} have been able to describe the data on elliptic flow very well.  The only points worthy of comments here are those outside the realm of what has been covered in the references given above.  There are then only three points:  (a) Is early thermalization necessary?  (b) What is the role of the shower partons?  (c) At what $p_T$ does the quark number scaling begin to break down?

\subsection{Effects of ridge formation at low $p_T$}

The usual hydrodynamical treatment of elliptic flow assumes rapid thermalization with initial time of expansion set at $\tau _0 = 0.6$ fm/c.  Such an early time of equilibration has never been shown to be the consequence of any dynamical process that is firmly grounded and commonly accepted.   The question then is whether the azimuthal anisotropy can be driven initially not by pressure gradient at $\tau < 1$ fm/c,  but by some other mechanism that is sensitive to the early spatial configuration.  That mechanism is suggested in Ref.\ [\refcite{5.11}] to be semihard scattering, the rate of which can be high for parton $k_T$ around 2-3 GeV/c, while the time scale involved is low ($\sim 0.1$ fm/c).  Semihard partons created near the surface of the nuclear overlap can lead to a continuous range of ridges that is shaped by the initial geometry.  The effects of such ridges are not considered in the usual studies in hydrodynamics, but should not be ignored.  It is shown in Refs.\ [\refcite{5.11,4.14}] that the azimuthal anisotropy of the ridges produces the observed $v _2$ at low $p_T$ without the use of hydrodynamics.  Elliptic flow at some point of the expansion of the system may well be describable by hydrodynamics, but early thermalization is not required.

For $dN/d\phi$ of single-particle distribution, no triggers are used and ridges are generated by many semihard partons produced in each event.  Semihard scatterings take place throughout the overlap, but only those occurring near the surface and directed normal to the surface can lead to the development of ridges, as discussed in Sec.\ 4.4.  If a drastic simplification is made to require ridge particles to be also directed only in the the directions normal to the almond-shaped initial boundary in the transverse plane, then the $\phi$  distribution of all ridges, $R(p_T, \phi)$, is restricted to the region $\phi \in {\cal R}$, which is a set of angles defined by
\begin{eqnarray}
|\phi| < \Phi (b)\, \qquad{\rm and}\,   \qquad |\pi - \phi| < \Phi (b)\quad  ,
\label{5.1}
\end{eqnarray}
where $\Phi (b) = \cos ^{-1}(b/2R_A)$.\cite{5.11}   Fluctuation from the restricted range is, of course, possible, not only because the semihard partons can have any scattered angle, but also because the ridge particles can fluctuate from the directions of the parton trajectories.  It is shown in the Appendix of Ref.\ [\refcite{4.14}] that the region ${\cal R}$ is nevertheless a good approximation even when all those effects plus elliptic geometry are taken into account, provided that the inaccuracy in the regions around $|\phi| \sim \pi/2$ for noncentral collisions is unimportant --- which is indeed the case, since the density at the upper and lower tips of the ellipse is low so ridge production there is suppressed.

Thus in the box approximation the ridge distribution is
\begin{eqnarray}
R(p_T, \phi) =  R(p_T) \Theta (\phi)    \ ,
\label{5.2}
\end{eqnarray}
where
\begin{eqnarray}
\Theta (\phi) = \theta (\Phi - |\phi|) + \theta (\Phi - |\pi - \phi|)    \ .
\label{5.3}
\end{eqnarray}
In the assumption that the above anisotropy from the ridge is the only $\phi$ dependence in $dN/d\phi$, replacing the usual assumption of rapid thermalization and the consequent pressure gradient, the bulk medium is then isotropic and the single-particle distribution at low $p_T$ can be written in the form
\begin{eqnarray}
{dN \over p_T dp_T d\phi}=  B(p_T) +   R(p_T) \Theta (\phi)  \ .
\label{5.4}
\end{eqnarray}
The normalized second harmonic in $\phi$ can then be calculated analytically, yielding
\begin{eqnarray}
v_2 (p_T,b) = \left<\cos 2 \phi\right >_b  = {\sin 2 \Phi (b)  \over \pi B(p_T)/R(p_T) + 2 \Phi (b)}  \ .
\label{5.5}
\end{eqnarray}
At low $p_T$ the first term in the denominator is much larger than the second, so
(\ref{5.5}) is reduced to the even simpler formula
\begin{eqnarray}
v_2 (p_T, b ) \simeq  { R(p_T)  \over \pi B(p_T)}  \sin 2 \Phi (b)  \ ,
\label{5.6}
\end{eqnarray}
This is such a compact formula that its validity should be checked regardless of its derivation for the sake of having available a simple description of what is usually called elliptic flow after extensive computation.

The $p_T$ distributions of $B(p_T)$ and  $R(p_T)$ are given, respectively, in (\ref{4.8}) and (\ref{4.9}).  The latter can be written in a form that factors out $B(p_T)$ so that $R/B$ has the simple form
\begin{eqnarray}
{R(p_T) \over B(p_T)}  =  e^{E_T(p_T)/T''_h}-1   \ ,
\label{5.7}
\end{eqnarray}
which becomes $E_T(p_T)/T''_h$ at low $p_T$, since $T''_h$ is large.  It then follows from (\ref{5.6}) that for small $E_T$
\begin{eqnarray}
v^h_2 (E_T, b) = {E_T \over \pi T''_h} \sin 2\Phi(b) \ .
\label{5.8}
\end{eqnarray}
Thus the initial slope in $E_T$ depends only on $T''_h$ (which sets the scale) and the geometric factor $\sin 2\Phi(b)$.  That factorizability is in agreement with the data for pion, but not so well for proton for mid-central to peripheral collisions.  That is because $T''_p$ develops a $b$ dependence when the centrality $c$ is above $0.3$, as can be seen in (\ref{4.7}).  While $T''_{\pi}$ is essentially constant $T''_p(c)$ can be approximated by a polynomial in $c$ \cite{4.14}
\begin{eqnarray}
T''_{\pi} = 1.7\ {\rm GeV},  \qquad T''_p= 2.37\,(1 - 1.05 c + 0.26 c^2)\ {\rm GeV} \ .
\label{5.9}
\end{eqnarray}
For $E_T \sim 1$ GeV the full expression in (\ref{5.7}) should be used in (\ref{5.6}), instead of the small $E_T$ approximation in (\ref{5.8}).  The results for $\pi$ and $p$ are shown in Fig.\ 9, and reproduce the data\cite{5.1} very well.  The reason why $T'_p(c)$ decreases with increasing peripherality, resulting in similar trend in  $T''_p(c)$, is that 3-quark recombination is more difficult when the thermal partons become less abundant at lower medium density as the collisions get more peripheral.  At fixed $p_T \sim 0.5$ GeV/c the $b$ dependences of $v^h_2$ for $h = \pi$ and $p$ are studied in Ref.\ [\refcite{4.14}] in which it is shown that the characteristics of the data are well captured by the simple formula in (\ref{5.5}).

\begin{figure}[th]
\vspace{-1cm}
\centerline{\psfig{file=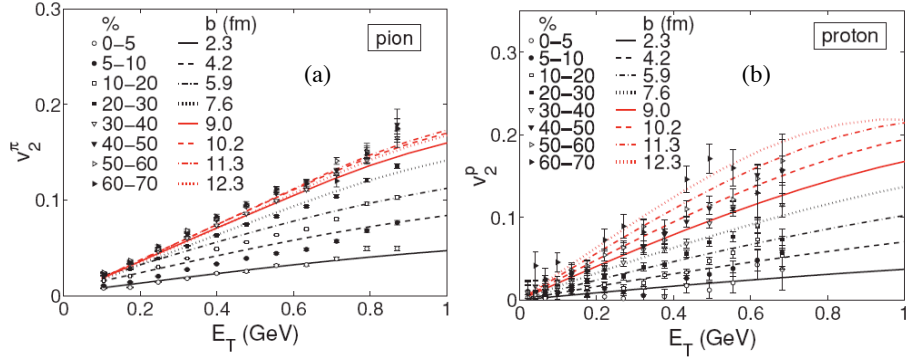,width=14.5cm}}
\vspace*{-5.8cm}
\caption{(a) Pion $v_2$ and (b) proton $v_2$, calculated in the RM\cite{4.14} by taking ridge effect into account. The data are from Ref.\ [\refcite{5.1}].}
\end{figure}

\subsection{Effects of shower partons at intermediate  $p_T$}

As $p_T$ is increased to above 2 GeV/c, it is necessary to consider the role played by the shower partons, which introduce $\phi$ dependence due to jet quenching of hard parton that depends on path length in the medium.  The dominance of TS + SS recombination over TT leads to a change in the $p_T$ dependence of $v_2(p_T)$.

The shower parton distribution ${\cal S}(q)$, given in (\ref{3.4a}), due to a hard parton produced at momentum $k$, is for central collisions and is averaged over all $\phi$.  Now for noncentral collisions with $\phi$ anisotropy, that formula needs to be generalized.  Assuming that the energy loss of a hard parton is proportional to $\sqrt{k'}$ where $k'$ is the initial parton momentum,\cite{1.32,5.12} one can write $\Delta k = k' - k$ in the form
\begin{eqnarray}
\Delta k =  \epsilon (b)  \hat{\ell} (b, \phi) \sqrt{k'} \ ,
\label{5.10}
\end{eqnarray}
where $ \hat{\ell} (b, \phi)$ is the normalized path length and $\epsilon (b)$ is the energy-loss coefficient that depends, apart from geometrical factors, a parameter $ \epsilon_0$ that is to be determined.  After solving  (\ref{5.10})  for $k'$, and replacing $\zeta f_i(k)$ in  (\ref{3.4a})  by $ f_i(k')$ at the shifted momentum, one can keep the first two non-vanishing terms in the harmonic expansion of $f_i(k'(k, b, \phi))$ and get
\begin{eqnarray}
f_i(k'(k, b, \phi)) = f_i (k)  \left[g_0(k, b) + 2 g_2(k, b) \cos 2 \phi \right] ,
\label{5.11}
\end{eqnarray}
where $g_0$ and $g_2$ can be determined explicitly in terms of $\epsilon _0$, $k$ and $b$.\cite{4.14}  The shower contribution to the single-pion distribution, when averaged over all $\phi$, may be written in the symbolic form
\begin{eqnarray}
{dN^{TS + SS}_{\pi}(b) \over p_Tdp_T} = {1 \over 2 \pi} g_0(k, b) f_i(k) \otimes ({\cal TS} + {\cal SS})
\label{5.12}
\end{eqnarray}
The contribution of the shower component to $v^{\pi}_2 (p_T, b)$ is
\begin{eqnarray}
v^{\pi, {\rm sh}}_2 (p_T, b) = {g_2(k, b) f_i (k) \otimes \left({\cal TS} + {\cal SS} \right) \over g_0(k, b) f_i (k) \otimes \left({\cal TS} + {\cal SS} \right)} .
\label{5.13}
\end{eqnarray}
The thermal component $v^{\pi, {\rm th}}_2 (p_T, b) $ is as given in (\ref{5.5}) and (\ref{5.7}).  The overall $v^{\pi}_2$ is obtained from the above with the help of an interpolating function $W(p_T, b)$
\begin{eqnarray}
 v ^{\pi}_2 (p_T, b) &=& v ^{\pi, \rm th}_2 (p_T, b) W(p_T, b)  +v ^{\pi, \rm sh}_2 (p_T, b) \left[1 - W(p_T, b)\right] \ .
\label{5.14}
\end{eqnarray}
where
\begin{eqnarray}
W (p_T, b) = {{\bf TT} \over {\bf TT}+{\bf TS} + {\bf SS}} \ ,
\label{5.15}
\end{eqnarray}
with ${\bf TT} $ being the thermal, and ${\bf TS} + {\bf SS}$ the shower, components of the $\phi$-averaged $dN_{\pi}/p_T dp_T$.  By fitting the single-pion distribution at 0-10\% centrality over the range $1 < p_T < 6$ GeV/c using all three ${\bf TT + TS + SS}$ terms, $\epsilon _0$ is found to be 0.55 GeV$^{1/2}$.  It is then possible to calculate $v ^{\pi}_2 (p_T, b)$ without any further adjustment; the result is shown in Fig.\ 10(a).  The data points for $E_T >1$ GeV are from Ref.\ [\refcite{5.13}].  The saturation of $v ^{\pi}_2$ in that range is thus interpreted in the RM as being due to the shower partons, where $W (p_T, b) $ is suppressed and $g_2$ is much smaller than $g_0$ at high $k$.

\begin{figure}[th]
\vspace{-1cm}
\centerline{\psfig{file=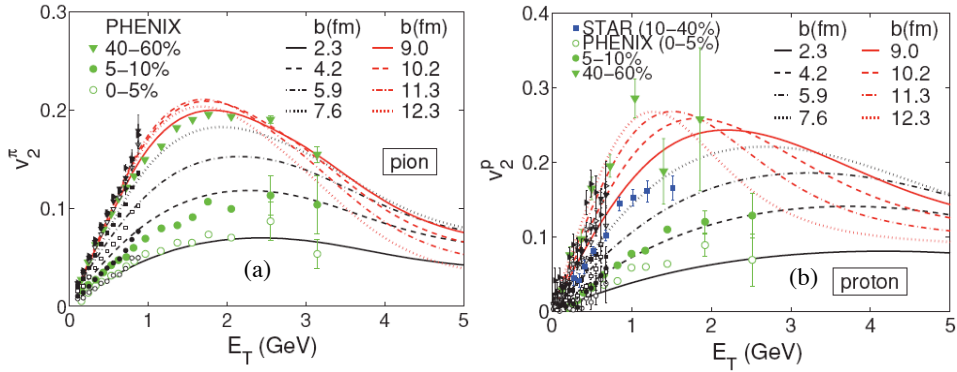,width=14.5cm}}
\vspace*{-5.6cm}
\caption{(a) Left panel: pion $v_2$ and (b) Right panel: proton $v_2$, both at higher $p_T$ where the effects of shower partons are taken into account. The data are Ref.\ [\refcite{5.3,5.13}].}
\end{figure}

For proton $v_2$ the general procedure in the calculation is similar to that for pion, except for an additional thermal parton to incorporate.  There is also the complication of $b$ dependence in the inverse slope $T'_p$ in (\ref{4.7}).  Taking them all into account the result for $v^p_2$ is shown in Fig.\ 10(b).  Comparison with data\cite{5.3,5.13} is acceptable,  although more accurate data are needed to check the calculated results at high $E_T$ and $b$.

What is learned from this study is that the main source of $\phi$ anisotropy is the path-length dependence of jet quenching, which is parametrized by one unknown $\epsilon _0$ that is determined by fitting the inclusive distribution at one value of $p_T$.  The characteristics of $v ^h_2 (p_T, b)$ in Fig.\ 10 are obtained without any more adjustable parameters.

It should be noted that we have refrained from using the term ``elliptic flow'', except in reference to past work based on hydrodynamics.  In Sec.\ 5.1 the emphasis is on the effect of ridges due to semihard scattering that are not taken into account by early-time hydrodynamics, although the exponential behavior of the thermal partons may well be the result of late-time hydrodynamical flow.  Then in this section hard scattering is incorporated in the treatment of jet-medium interaction through TS recombination before fragmentation dominates.  Such interaction is outside hydrodynamics, so the overall characteristics of $v ^h_2 (p_T, b)$ are not the properties of flow.  In Ref.\ [\refcite{1.32}] the effects of jet quenching on the hard partons are also considered, but since only the fragmentation of those partons is included,  the transitional contribution from TS interaction is not explicitly taken into account.  That turns out to be important in the intermediate $p_T$ region.

\subsection{Breaking of quark number scaling}

Quark number scaling (QNS) has long been regarded as a signature of recombination\cite{1.31,1.32,5.9} and has been verified in a number of experiments.\cite{5.1,5.3,5.4}  It has been regarded as a statement of the universality of $v ^h_2 (p_T/n_q)/n_q$, where $n_q$ is the number of constituent quarks in the hadron $h$.  It is based on the assumption of factorizability of the distribution of the quarks that recombine, i.e., the multiquark distributions in (\ref{3.1}) and (\ref{3.2}) (but with $\phi_i$ dependencies included) can be written as
\begin{eqnarray}
F_{n_q}\left(q_1, \phi_1;q_2,\phi_2; \cdots \right)=\prod^{n_q}_{i = 1} F_i (q_i,\phi_i)
=\prod^{n_q}_{i = 1} F_i (q_i) \left[1 + 2v^i_2 (q_i) \cos 2 \phi _i \right].
\label{5.16}
\end{eqnarray}
Coupled with the assumption that the RF has the simple form $\delta (q_i - p_T/n_q)$, it then follows trivially from (\ref{3.1}) and (\ref{3.2}) that
\begin{eqnarray}
v^h_2 (p_T) = n_q v^q_2 (p_T/n_q) \ ,
\label{5.17}
\end{eqnarray}
if the $v^i_2 (q_i)$ of all quarks in (\ref{5.16}) are the same, denoted by $v^q_2 (q)$.

From our discussions throughout this paper it is clear that none of the above assumptions are valid under close examination.    TT+TS+SS for pion and TTT+TTS+TSS+SSS for proton are obviously not factorizable.  Even at low $p_T$ where only the recombination of thermal partons is important, the inverse slopes $T'_h$, given in
(\ref{4.6}) and (\ref{4.7}), are not the same for $\pi$ and $p$.  Consequently, the $R/B$ ratios for $\pi$ and $p$ are different, as seen in (\ref{5.7}) and (\ref{5.9}), resulting in different $v^h_2 (E_T,b)$.  Furthermore, the wave functions of $\pi$ and $p$ are very different, since the pion is a tightly bounds state of the constituent quarks, while the proton is loosely bound.  That means the momentum fractions of the quarks (valons) are not $1/2$ for pion and $1/3$ for proton.  It is then a very rough approximation to write the momentum conservation $\delta$-function, $\delta (\Sigma q_i - p_T)$, as $\delta (n_q q_T- p_T)$ with a common $q_T$.  At intermediate $p_T$ where shower partons become important, we have seen that they acquire the $\phi$ dependence of the hard parton, given in (\ref{5.11}), so $v^S_2 (q_i)$ for the shower is different from $v^T_2 (q_j)$ for the thermal parton.  Even if TS and TTS contributions dominate, one can at best, by ignoring all other complications, have
\begin{eqnarray}
v^M_2 (p_T) = v^T_2 (q_1) + v^S_2 (q_2), \qquad v^B_2 (p_T) = v^T_2 (q_1)+v^T_2 (q_2)+v^S_2 (q_3) \  .
\label{5.18}
\end{eqnarray}
They would not lead to QNS, as expressed in (\ref{5.17}).

Most data in support of QNS are for minimum bias and at low $E_T$.  It is shown in Ref.\ [\refcite{4.14}] that the calculated result at $E_T/n_q < 0.5 $ GeV does exhibit QNS in agreement with the data, but the scaling is broken above that.  The breaking of QNS is due primarily to the nonequivalence of the $\phi$ dependencies of the thermal and shower partons.  The important point to stress here is that the breaking of QNS at intermediate $E_T$ does not imply the failure of recombination (in fact, it is expected), but the validity of QNS at lower $E_T$ does confirm recombination as the proper hadronization process.

\section{Hadron Correlation in Dijet Production}

In Sec.\ 4 the subject of discussion is ridgeology, which is the study of the effects of jets on the medium.  In this section we consider the reverse, i.e., the effects of the medium on the jets for $p_T$ not so extremely high as to exclude TS recombination.  In triggered events the ridge is the broad pedestal on top of which sits the peak.  The structure of that peak, when compared to the jet structure in $pp$ collisions, reveals the medium effect on hard partons.  Since the peak is restricted to a narrow cone around the trigger direction, it is natural in the RM to associate it with shower partons, while the ridge being broad is associated with the thermal partons.  Hadrons closely correlated to a trigger particle as jets exhibit peaks on both the same and away sides.\cite{6.1}-\cite{6.5}  How the structures of the jets on the two opposite sides differ from each other is a strong indication of the difference in energy losses in the two jets, since their path lengths in the medium are generally different.  In realistic collisions even when centrality is chosen to be within a narrow range, the path lengths can vary significantly depending on the location and angle of a scattered hard parton.  Thus a careful study of the properties of the near- and away-side jets must start with finding a good description of the variation of energy loss within each class of centrality.

\subsection{Distribution of dynamical path length}

By dynamical path length we mean not only the geometrical length of a trajectory, but also the medium effect along that trajectory.  To quantify that in an analytic expression, we need to revisit the single-particle distribution discussed earlier for central collision, but now formulated in a way appropriate for any centrality.  For pion production we have for $y \sim 0$
\begin{eqnarray}
{dN_{\pi} \over pdp} &=& {C^2 \over 6} e ^{-p/T}
+ {1 \over p^2} \sum_i \int {dq \over q} F_i (q)  \left[\widehat{\sf TS} (q, p)+{p \over q} D^{\pi}_i \left({p\over q} \right)\right] \quad ,
\label{6.1}
\end{eqnarray}
where all momenta are in the transverse plane with the subscript $T$ omitted.  The first term on the right side is from (\ref{4.3}) for TT recombination; the centrality dependence of $C$ is given in Ref.\ [\refcite{6.5a}].   The first term in the square bracket is for TS recombination which will be detailed below, and the second term is the fragmentation function that is equivalent to SS recombination.  $F_i(q)$ is the distribution of parton $i$ at the surface of the medium with $q$ denoting the momentum of the hard parton there.  It differs from the distribution $f_i(k)$ given in (\ref{3.4b}), that describes the hard parton with momentum $k$ at the point of creation.  In Secs.\ 2 and 3 where only central collisions are considered, an average suppression factor $\zeta$ is used, as shown in (\ref{3.4a}).  We now replace $\zeta f_i(k)$ by a path-dependent term related to $F_i(q)$ by
\begin{eqnarray}
F_i (q) = \int^L_0 {dt \over L} \int dk k f_i (k) G(q, k, t) ,
\label{6.2}
\end{eqnarray}
where $G(q, k, t)$ is a degradation factor that describes the decrease of parton momentum from $k$ to $q$ as the hard parton traverses a distance $t$ through the medium.\cite{6.6}  $L$ is the average maximum length of that trajectory.  In the limit $L \to 0$, $F_i(q)$ should become the parton distribution function $F_i(k)$ for $pp$ collisions.

For energy loss we seek a form that is consistent, on one hand, with $\Delta E \propto L$ as suggested in Ref.\ [\refcite{4.2}] for 1D expansion, and on the other hand, with $\langle dE/dL\rangle  \propto E$ in Ref.\ [\refcite{6.8}] for $6 < E < 12$ GeV.  A reasonable approximation of the differential energy loss is then
 \begin{eqnarray}
{\Delta E\over E}=\beta \Delta L  ,
\label{6.3}
\end{eqnarray}
which translates to our variables as
\begin{eqnarray}
k-q=k\beta t .
\label{6.4}
\end{eqnarray}
at small $t$ with $\beta$ being an adjustable parameter.  For larger $t$ we exponentiate the above to
\begin{eqnarray}
q=k e^{-\beta t} \quad ,
 \label{6.5}
\end{eqnarray}
and let $G(q, k, t)$ take the simple form
\begin{eqnarray}
G(q, k, t) = q \delta(q - k e^{-\beta t})  \quad .
\label{6.6}
\end{eqnarray}
The $\delta$ function can be broadened to account for fluctuations, but we shall take (\ref{6.6}) as an adequate approximation of the complicated processes involved in the parton-medium interaction, the justification of which rests ultimately on how well the calculated result can agree with the data for $2<p_T<12$ GeV/c.

Using (\ref{6.6}) in (\ref{6.2}) yields
\begin{eqnarray}
F_i (q) = {1 \over \beta L}\int^{q e^{\beta L}}_{q} dk k f_i (k) \quad ,
\label{6.7}
\end{eqnarray}
which shows explicitly how $f_i(k)$ is transformed to $F_i(q)$ by the nuclear effect parametrized by $\beta L$, while $f_i(k)$ itself contains the hidden modifications due to such effects as $N_{\rm coll}$ dependence and shadowing.\cite{3.7a}  Since $f_i(k)$ is $dN^{\rm hard}_i/kdkdy|_{y = 0}$, (\ref{6.7}) becomes in the limit of $L \to 0$ the invariant distribution for hard parton $i$ production in $pp$ collision.  In heavy-ion collisions Eq.\ (\ref{3.4a}), written for centrality $\sim 0\%$, can now be improved to the form
\begin{eqnarray}
{\cal S}(q_1) = \int {dq \over q} F_i (q) S^{j}_i (q_1/q)
\label{6.8}
\end{eqnarray}
for any centrality with $L$ being the geometrical path length.  We can now write the TS recombination term in (\ref{6.1}) as
\begin{eqnarray}
\widehat{\sf TS} (q, p) = \int {dq_1 \over q_1} S^j_i \left({q_1 \over q}\right) \int dq_2 C_{\bar j} e^{-q_2/T'} {R}_{j \bar{j}}(q_1, q_2, p) ,
\label{6.9}
\end{eqnarray}
where $R_{j\bar{j}}$ is the RF given in (\ref{4.2}).

Eq.\ (\ref{6.1}) is now totally specified, relating the observable pion spectrum to the nuclear parameter $\beta L$ through $F_i(q)$ given in (\ref{6.7}).  We emphasize that because all three components of TT+TS+SS recombination are included in (\ref{6.1}) it can describe the pion distribution at all $p_T$ using only one parameter $\beta L$ for each centrality.  Fig.\ 11(a) shows the fits of the $\pi ^0$ distributions  for nine bins of centrality,\cite{6.9} with a different value of $\beta L$ used for each centrality.\cite{6.6}  Those values of $\beta L$ are shown by the nine points in Fig.\ 11(b) (ignoring the line for the moment).  Evidently, the agreement with data is excellent over such a wide range of $p_T$ and centrality $c$.  Apart from concluding that the RM works well, the jet-medium interaction has been effectively summarized by one phenomenological function $\beta L(c)$, which could not have been extracted from the data without a reliable way to relate energy loss to the suppression of pion production at all $p_T$ through appropriate hadronization.

\begin{figure}[th]
\vspace{-1cm}
\centerline{\psfig{file=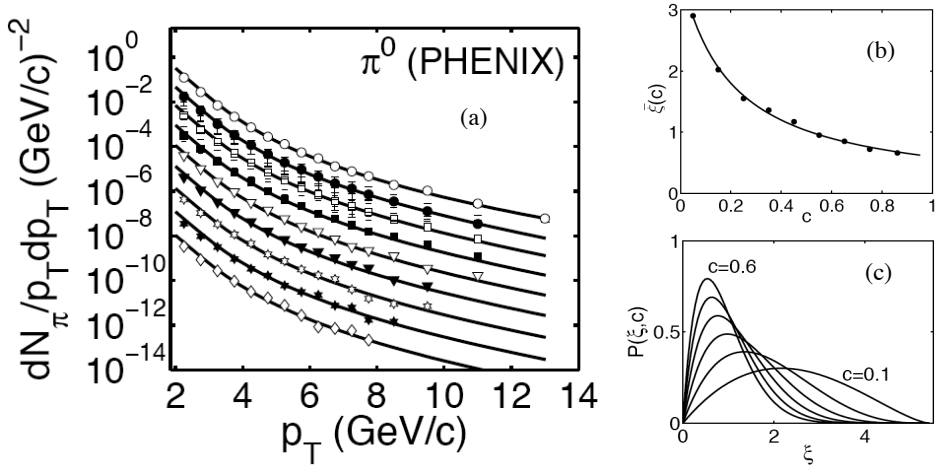,width=14.5cm}}
\vspace*{-4.2cm}
\caption{(a) Pion spectra\cite{6.9} for various centralities lowered by a factor of 0.2 for each step of increase of 10\%. The solid lines are fits in the RM\cite{6.6} with one parameter for each centrality $c$. Those parameters are shown by dots in (b), which are fitted by the curve  $\bar\xi(c)$ that is the average of the dynamical path length $\xi$ over $P(\xi,c)$ defined in Eq. (\ref{6.11}). (c) $\xi$ distributions for various $c$.}
\end{figure}

Having obtained  $\beta L(c)$ we now go a step further to inquire what kind of variation of the dynamical path length can the nuclear overlap generate for any given centrality.  That is, for a fixed $c$, $\beta L(c)$ is the average of a variable $\xi$ over a probability distribution $P(\xi, c)$ that describes the likelihood that a particular trajectory occurs at centrality $c$ with an effective energy loss such that
\begin{eqnarray}
\beta L (c) = \int d \xi \xi P (\xi, c)  .
\label{6.10}
\end{eqnarray}
Thus $\xi$ plays the role of $\beta L$ except that it can vary from $0$ to a maximum for every fixed initial elliptic geometry depending on the initial point and orientation of the hard parton.  So $\xi$ is the dynamical path length of a trajectory whose average $\bar{\xi}(c)$ is $\beta L (c)$.  In Ref.\ [\refcite{6.6}] $P (\xi, c)$ is chosen to have the form
\begin{eqnarray}
P(\xi, c)=N\xi(\xi_0-\xi)^{\alpha c},
\label{6.11}
\end{eqnarray}
where $N$ normalizes the total probability to $1$.  The two parameters $\xi_0$ and $\alpha$ are adjusted to fit the nine points in Fig.\ 11(b), with the result shown by the curve that renders an excellent fit for
\begin{eqnarray}
\xi_0=5.42,   \qquad  \alpha=15.2    \ .
\label{6.12}
\end{eqnarray}
Thus  (\ref{6.11})  is a very efficient way to describe the energy loss effect for all centralities.  The shapes of $P (\xi, c)$ are shown in Fig.\ 11(c) for 6 values of $c$, exhibiting the expected peak that decreases with increasing $c$ or shrinking ellipse.  In view of the difficulty of deriving $\beta L (c)$ from first principles, let alone $P(\xi, c)$, it is very convenient to have the path-dependent quenching effect be represented by the simple description in (\ref{6.11}) and (\ref{6.12}).

The single-pion distribution at midrapidity for centrality $c$ can now be expressed as
\begin{eqnarray}
{dN_{\pi}(c)\over pdp} = \int d\xi P(\xi,c){dN_{\pi}(c,\xi)\over pdp},
 \label{6.13}
\end{eqnarray}
where (\ref{6.1}) is to be used for $dN_{\pi}(c,\xi)/pdp$, provided that $F_i(q)$ in it is not as given in  (\ref{6.7}), but with $\beta L$ replaced by $\xi$, and, of course, $f_i(q)$ scaled by $N_{\rm coll}(c)$.  It should be clear that (\ref{6.1}) has basically two parts:  $F_i(q)$ that describes the hard parton part and the rest that describes the hadronization part.  We have in this subsection incorporated the centrality-dependent energy-loss effect on $F_i(q)$ by the use of just two dimensionless parameters given in (\ref{6.12}).  Having successfully formulated the treatment of the single-hadron distribution, we are now ready to proceed to the study of dihadron correlation in near- 
and away-side jets.

\subsection{Near-side and away-side yields per trigger}

For dihadron correlation many momentum vectors of partons and hadrons are involved.  To depict clearly their relationships with one another, we show in Fig.\ 12 a pictorial representation of all of them.  The near side is on the right and the away side on the left.  The vectors $k$, $q$, and $p_t$ are, respectively, the momenta of the initiating hard parton, of the same parton as it leaves the medium, and of the trigger hadron.  The associated hadron on the same side is labeled by $p_a$.  On the away side the  corresponding momenta are $k'$, $q'$, and $p_b$, there being no trigger on that side.  If there are no transverse momenta in the beam partons, $k$ and $k'$ should be equal and opposite in every event; however, their averages over triggered events may not be the same, so it is better to label them distinctly from the start. The momenta of interest in the following are such that the hadronic $p_T$ are in the range $2<p_T<10$ GeV/c but with special emphasis on $p_T<6$ GeV/c.

\begin{figure}[th]
\vspace{-1cm}
\centerline{\psfig{file=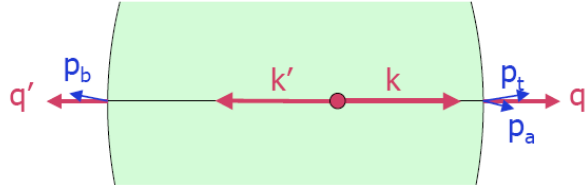,width=16cm}}
\vspace*{-8.4cm}
\caption{A sketch of momentum vectors of partons (in red) and hadrons (in blue) with near side being on the right and away side on the left.}
\end{figure}

For two pions in the same jet, neither of which are in the ridge, we can leave out the TT contribution, and write as a generalization of (\ref{6.1})
\begin{eqnarray}
{dN_{\pi\pi}(\xi) \over p_tp_a dp_t dp_a}&=& {1 \over (p_tp_a)^2} \sum_i \int {dq \over q} F_i (q, \xi)  \left\{\left[\widehat{\sf TS}  (q,p_t) + {p_t \over q} D^{\pi}_i  \left({p_t \over q}\right)\right]\widehat{\sf TS}  (q - p_t,p_a) \right.  \nonumber\\
&&+  \widehat{\sf TS}  (q - p_a,p_t){p_a \over q}D^{\pi}_i  \left({p_a \over q}\right)
\left. +  {p_t p_a \over q^2_j }D^{\pi}_2  \left( {p_t \over q},  {p_a \over q}\right)\right\}
\label{6.14}
\end{eqnarray}
where $D_2(z_1, z_2)$ is the dihadron fragmentation function.\cite{6.6}  For notational brevity (\ref{6.14}) is for a fixed $\xi$, the averaging over $P(\xi, c)$ being a process that can be applied when $c$ is fixed.  For two pions on opposite sides the recoil parton must be considered explicitly, so (\ref{6.2}) should be generalized to
\begin{eqnarray}
F'_i (q,q',\xi) &=& \int ^{\xi}_0 {\beta dt \over \xi} \int dk k f_i (k) G (q, k, t) G (q', k',{\xi \over \beta}- t)\nonumber\\
&&= {1 \over \xi} \int^{qe^{\xi}}_{q} dk k f_i (k) q q' \delta(q q' - k k' e^{- \xi}) .
\label{6.15}
\end{eqnarray}
with $k' = k$.  Thus the dipion distribution for $p_b$ on the side away from $p_t$ is
\begin{eqnarray}
{dN_{\pi\pi}(\xi) \over p_t p_b dp_t dp_b} &=& {1 \over (p_t p_b)^2} \sum_i \int {dq \over q} {dq' \over q'}F'_i (q,q',\xi) \left[\widehat{\sf TS}  (q,p_t) + {p_t \over q} D^{\pi}_i\left({p_t \over q}\right)\right]\nonumber\\
&& \times  \left[  \widehat{\sf TS}  (q', p_b) + {p_b \over q'}D^{\pi}_{i'}\left({p_b \over q'}\right)   \right] .
\label{6.16}
\end{eqnarray}
The $\delta$ function in (\ref{6.15}) restricts the integration of $q'$ in (\ref{6.16}) to the range from $qe^{- \xi}$ to $qe^{\xi}$ that correspond to the hard scattering point being on the near-side boundary to the far-side boundary.

The yield per trigger at fixed centrality can now be obtained for the near side as
\begin{eqnarray}
Y^{\rm near}_{\pi\pi} (p_t, p_a, c)= {\int d \xi P (\xi, c) dN_{\pi\pi}(\xi)/p_t p_a dp_t dp_a \over \int d \xi P (\xi, c) dN_{\pi}(\xi)/p_t dp_t}  ,
\label{6.17}
\end{eqnarray}
and similarly for $Y^{\rm away}_{\pi\pi} (p_t, p_b, c)$ with $dN_{\pi\pi}(\xi)/p_t p_b dp_t d p_b$ being used in the numerator.  Since every factor is already specified, it remains only for the computation to be carried out.  The results are shown in Fig.\ 13(a) for near side and (b) for away side.  The three sheets for $c = 0.05$, $0.35$, and $0.86$ are separated by a factor of $10$ between sheets for clarity's sake.
Superficially, the two figures may look similar in a vertical scale that spans over 4 orders of magnitude, but there are significant differences that only by closer examination can one learn from them the nature of the medium effects on the jets.

\begin{figure}[th]
\vspace{-1cm}
\centerline{\psfig{file=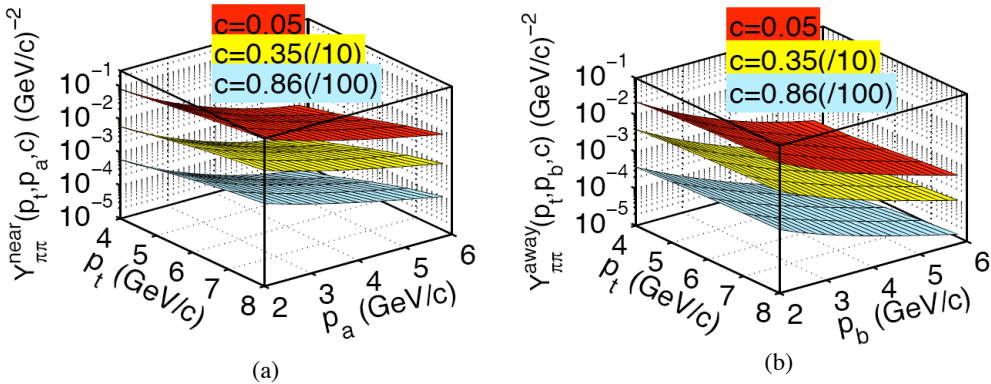,width=14.5cm}}
\vspace*{-4.8cm}
\caption{(a) Near-side jet yield per trigger vs trigger momentum $p_t$ and associated particle momentum $p_a$. (b) Away-side jet yield per trigger vs $p_t$ and $p_b$.}
\end{figure}

\subsection{Medium effects on dijets}

Let us first cut the two figures in Fig.\ 13 by three fixed-$p_t$ planes at $p_t = 4$, $6$, and $8$ GeV/c, and show the results in Fig.\ 14(a) and (b) for $c = 0.05$ and $0.35$.  Note (i) the dependence of $Y^{\rm near}_{\pi\pi} (p_a)$ on $p_a$ is more sensitive to $p_t$ than that of the dependence of $Y^{\rm away}_{\pi\pi} (p_b)$ on $p_b$; (ii) the increase of the yield with $p_t$ is more pronounced for the near side than for the away side; and (iii) $Y^{\rm near}_{\pi\pi}$ has negligible dependence on $c$, but $Y^{\rm away}_{\pi\pi}$ increases by roughly a factor of 2 when $c$ changes from $0.05$ to $0.35$.  Let us discuss these three features separately.

\begin{figure}[th]
\vspace{-1cm}
\centerline{\psfig{file=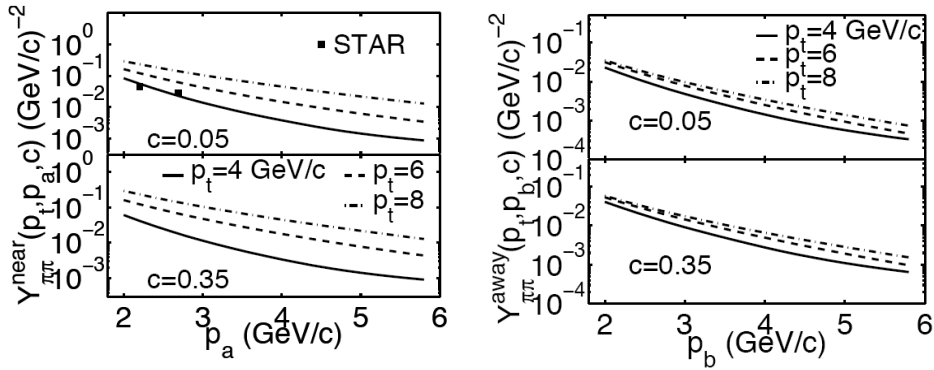,width=14.5cm}}
\vspace*{-5.8cm}
\caption{(a) Left panel: near-side jet yield per trigger vs $p_a$ for fixed $p_t$ and $c$. Data points are from Ref.\ [{6.11}]. (b) Right panel: the same for away side vs $p_b$.}
\end{figure}

On item (i) one can determine the near-side average inverse slope $T_a$ of the approximate exponential behavior in the range $2 < p_a < 4$ GeV/c for $c = 0.05$, for which there are data.  The result is shown by the line in Fig.\ 15(a), and agrees well with the data.\cite{3.5,4.7,6.10}  Evidently, the spectrum of the associated particles in a jet becomes harder as the trigger momentum increases, as one would expect.  It should be remarked that the line in Fig.\ 15(a) is for pions, while the data are for all charged particles. However, since pions dominate in jet peaks (unlike the ridges), the comparison is not unreasonable.  The inverse slope $T_b$ for the away-side jet at $c = 0.05$ for the same ranges of $p_t$ and $p_b$ is shown in Fig.\ 15(b), exhibiting a lower value compared to $T_a$ and a mild decrease with $p_t$.  There exist no suitable data for comparison.  The strong difference between $T_a$ and $T_b$ has good reasons, as will be discussed below.  Two data points from Ref.\ [\refcite{6.11}] are included in Fig.\ 14(a), the details of which are discussed in Ref.\ [\refcite{6.6}].  They lend support to the theoretical curve in both magnitude and $p_a$ dependences.

\begin{figure}[th]
\vspace{-1cm}
\centerline{\psfig{file=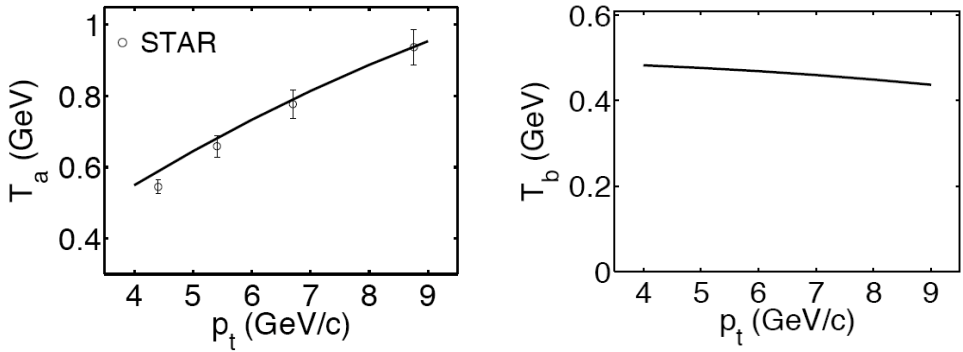,width=14.5cm}}
\vspace*{-6cm}
\caption{Inverse slopes of associated particles on the (a) near side and (b) away side. Data in (a) are from Ref.\ [\refcite{4.7,6.10}].}
\end{figure}

Note (ii) about the $p_t$ dependence of the yield is related to note (i) about the $p_t$ dependencies of the shapes in $p_a$ and $p_b$.  First of all, $Y^{\rm near}_{\pi\pi}$ is larger than $Y^{\rm away}_{\pi\pi}$ in magnitude, meaning that there is more suppression on the away side than on the near side.  To quantify that interpretation let $\xi$ be fixed at $2.9$ corresponding to the maximum probability for $c = 0.05$ so that the suppression effect is not partially hidden by averaging over $\xi$, which can vary from $0$ to $\xi_0$.  Let the suppression factor for the near side be defined by
\begin{eqnarray}
\Gamma_{\rm near}(p_T)=\langle e^{-\beta t}\rangle_{p_T} = \langle q/k \rangle_{p_T},
\label{6.18}
\end{eqnarray}
where the average is performed over $dN_{\pi}/p_T dp_T$ given in (\ref{6.1}), and the last expression follows from (\ref{6.6}).  $\Gamma_{\rm near}(p_T)$ gives a measure of the fraction of momentum retained after energy loss reduces $k$ to $q$.  For the away side the suppression factor is defined by
\begin{eqnarray}
\Gamma_{\rm away}(p_t, p_b)=\langle e^{-\xi + \beta t}\rangle_{p_t, p_b}
\label{6.19}
\end{eqnarray}
where $-\xi + \beta t$ is equivalent to $-\beta(L-t)$ if $\xi$ is denoted by $\beta L$ for a fixed medium length $L$, in which $t$ is the portion from the hard-scattering point to the near-side surface, and $L - t$ is the distance to the away-side surface.  The average in (\ref{6.19}) is done using the dihadron distribution given in (\ref{6.16}).

\begin{figure}[th]
\vspace{-1cm}
\centerline{\psfig{file=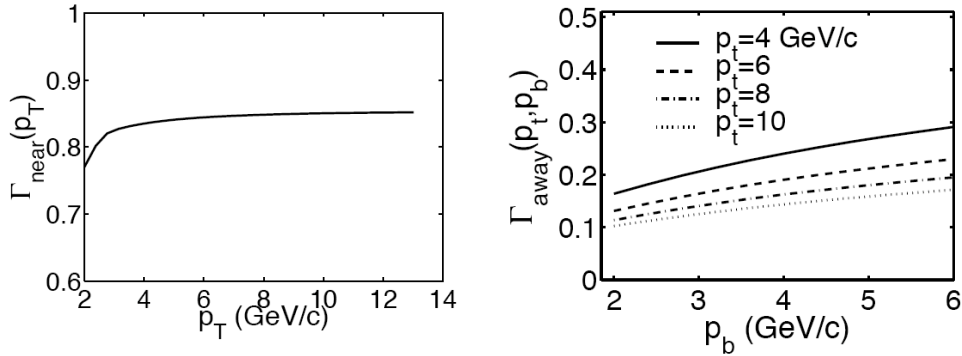,width=14.5cm}}
\vspace*{-5.9cm}
\caption{Suppression factors for the (a) near side and (b) away side.}
\end{figure}

The calculated results for $\Gamma_{\rm near}(p_T)$ and $\Gamma_{\rm away}(p_t, p_b)$ are shown in Fig.\ 16(a) and (b).  They clearly indicate that $\Gamma_{\rm near}(p_T)$ is much larger than $\Gamma_{\rm away}(p_t, p_b)$  with the implication that there is less suppression on the near side than on the away side.  The physics of the phenomenon is clear:  at large $p_T$ the point of creation of hard parton with large $k$ is predominantly close to the near-side surface in order to minimize energy loss with the consequence that the distance to the away-side surface is longer, thus more suppression for $p_b$ on that side.  Since  $\Gamma_{\rm near}(p_T)$ saturates at around 0.85, only 15\% of the parton energy is lost to the medium on the near side.  The corresponding $\langle \beta t \rangle$ is less than 0.2, so $\langle \beta t \rangle /\xi = \langle t \rangle/L \approx 0.065$, meaning that the hard partons are created within a layer of thickness $\sim 13\%$ of $L$ from the surface.  That is a quantitative description of trigger bias.  On the other hand, the behavior of $\Gamma_{\rm away}(p_t,p_b)$ reveals the opposite:  at fixed $p_b$ it decreases with increasing $p_t$, implying more suppression as the hard-scattering point is pulled closer to the near-side surface.  That is antitrigger bias.  At fixed $p_t$, $\Gamma_{\rm away}(p_t,p_b)$ increases with $p_b$, since higher $p_b$ demands higher $q'$, which can be satisfied only if $\langle L - t \rangle$ is reduced or $\langle k' \rangle$ increased, actually both.  Lowering $\langle L - t \rangle$ is, of course, a way to reduce energy loss by pulling the scattering point closer to the away-side surface.  But the increase of $\langle k' \rangle$ with $p_b$ is another aspect of antitrigger bias, whose details are described in Ref.\ [\refcite{6.6}].  Event-by-event momentum conservation requires $k' = k$; however, when averaged over all events, $\langle k' \rangle$ depends on $p_b$, while $\langle k  \rangle$ does not.  In general, $\langle k' \rangle$ is far greater than $\langle k \rangle$ because any trigger favors shorter $\langle t \rangle$, and any finite $p_b$ pushes up $\langle k' \rangle/\langle k \rangle$ though not $\langle k'/k \rangle$.  Because of the difficulty of producing an associated particle on the away side relative to one on the near side, $T_b$ is lower and decreases with increasing $p_t$ in Fig.\ 15(b), while $T_a$ is higher and increases with $p_t$ in Fig.\ 15(a).

Now on note (iii) about centrality dependence Fig.\ 14(b) shows that the dependencies of $Y^{\rm away}_{\pi \pi}(p_t, p_b, c)$ on $p_t$ and $p_b$ are essentially the same whether $c = 0.05$ or $0.35$, except that the magnitude of the yield is increased due to the reduction of path length at higher $c$.  For the near side there is essentially no dependence of $Y^{\rm near}_{\pi \pi}(p_t, p_a, c)$ on $c$.  That is shown more explicitly in Fig.\ 17, where the lower three lines are for $p_t = 4$ GeV/c and the upper line is for $p_t = 6$ GeV/c.  The solid lines are for integrated yields with $2 < p_a < 4$ GeV/c.  The near independence on $c$ is a manifestation of the trigger bias, since the hard-parton production point, being restricted to a layer roughly 13\% of $L$ just inside the near-side surface, is insensitive to how large the main body of the medium is.  Actually, the decrease of the TS component with $c$ balances the increase of the SS component with $c$ so that the net yield being their sum is approximately constant in $c$.  The data points in Fig.\ 17 support the calculated result in both the magnitude and the $c$ dependence of $Y^{\rm near}_{\pi \pi}(p_t, c)$ when $p_a$ is integrated from 2 to 4 GeV/c.\cite{4.7}

\begin{figure}[th]
\vspace{-1cm}
\centerline{\psfig{file=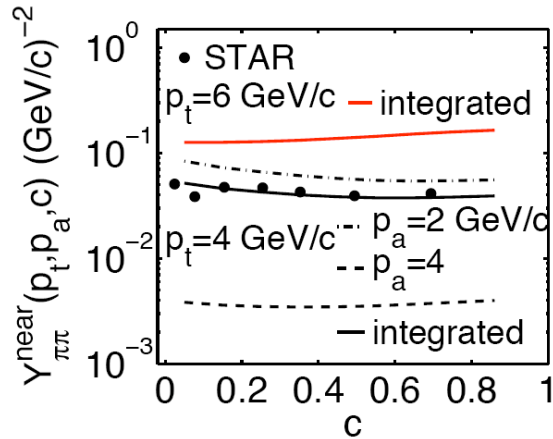,width=13.5cm}}
\vspace*{-3.9cm}
\caption{Near-side yield per trigger vs centrality $c$. The solid lines are for integrated yields with $2<p_a<4$ GeV/c. The data points are from Ref.\ [\refcite{4.7}].}
\end{figure}

In summary the discussion above gives quantitative demonstration that the trigger bias is the preference for the hard-scattering point to be close to the near-side surface and that the antitrigger bias is the consequence: $\langle k' \rangle$ is much larger than $\langle k \rangle$, and even larger than $p_t$ or $p_b$.  Those are the properties of the events selected by a trigger at $p_t$ with an associated particle on either the near or away side.

\subsection{Symmetric dijets and tangential jets}

We now follow the summary comment at the end of Sec.\ 6.3 with the question on what if we select events with symmetric dijets where $p_t = p_b$.  Instead of studying the properties of a third particle in association with the two trigger particles (for which we need trihadron correlation function), we can nevertheless learn a great deal from examining closely various calculable quantities in the dihadron correlation problem.  Let $p$ be the momentum of the symmetric dijets, $p = p_t = p_b$.  One can calculate $\langle k' \rangle(p, c)$ and $\langle q' \rangle (p, c)$ for various centralities, even though they are not directly measurable.  The result is that they both increase almost linearly with $p$ and that there is essentially no dependence on $c$.  Their ratio $\langle q' \rangle/\langle k' \rangle$ is therefore approximately constant in $p$ with a value of about 0.8 as shown in Fig.\ 18(a).  That behavior is similar to the property of $\Gamma_{\rm near}(p)$, shown in Fig.\ 16(a), which, according to (\ref{6.18}), is also $\langle q/k  \rangle$.  It is important to bear in mind that $\langle q' \rangle$ and $\langle k' \rangle$ are averages over the two-particle distribution $dN_{\pi \pi} (c) / p_tp_bdp_tdp_b$ with $p_t = p_b$, given in (\ref{6.16}) for fixed $\xi$, followed by averaging over $P(\xi, c)$, whereas $\langle q \rangle$, $\langle k  \rangle$ and  $\langle q/k \rangle$ are averages over the single-particle distribution $dN_{\pi}/pdp$, as noted after (\ref{6.18}).  The near-side averages know nothing about the away-side analysis, so $\Gamma _{\rm near} (p)$ describes only the suppression associated with trigger bias.    The fact that the suppression on the away side  is about the same as on the near side, when the average $\langle q' / k' \rangle$ is over symmetric dijets and $\langle q / k \rangle$ is averaged over near-side jet, and that it is true for any centrality, has only one important implication:  the dijets are created very near the surface on both sides.  It means that the symmetric dijets are dominated by tangential jets, which behave similarly at all centralities.  That is a striking conclusion.  Hard scattering can, of course, occur anywhere in the overlap region. But those partons from the interior lose most of their momenta on the way out.  Those that are created on the near side would not give rise to symmetric dijets.  Thus only those created along the rim and directed tangentially to the surface can lead to events with $p_t = p_b$.

\begin{figure}[th]
\vspace{-1cm}
\centerline{\psfig{file=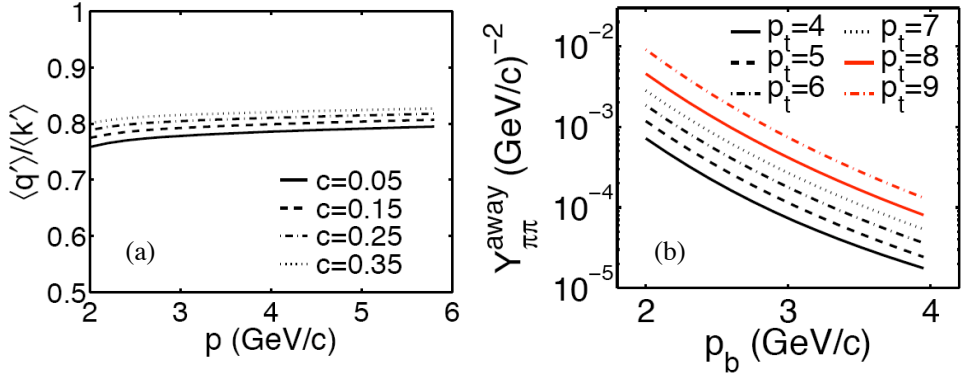,width=14.5cm}}
\vspace*{-5.5cm}
\caption{(a) The ratio $\langle q'\rangle/\langle k'\rangle$ at the symmetry point $p=p_t=p_b$ for four values of centrality. (b) Distribution of associated pion ($p_b$) in the away-side jet for six values of pion trigger momentum ($p_t$) in GeV/c for $\xi$ fixed at 2.9.}
\end{figure}

There is some experimental evidence to support this finding.  In 2jet + 1 correlation studied\cite{6.12} it is found that the third particle does not show any ridge structure and that the centrality dependence goes as $N^{2/3}_{\rm part}$.  The latter means that the dijets are created near the surface, while the former means that the dijets are tangential because we have already seen in Sec.\ 4.4 that ridge production depends on the local flow direction to match the trigger direction due to the correlation function in (\ref{4.13}).  The flow direction near the surface is normal to the surface and is therefore normal to tangential jets.  In the experimental analysis\cite{6.12} the two  jets have $p_T$ cuts at $p_{T_1}> 5$ GeV/c and  $p_{T_2}> 4$ GeV/c, so they are not exactly symmetrical.   If ridge formation from those dijets is to be discovered in the future, it would pose a serious challenge to the treatment of ridgeology in Sec.\ 4.  But so far the data are in accord with the results both here and on ridges.

\subsection{Unsymmetric dijets and tomography}

The conclusion in the last  section on the dominance in symmetric dijets by tangential jets leads one to consider the only option left for probing the dense medium by parton jets apart from using direct $\gamma$, and that is the study of away-side jets in unsymmetric dijets.  If hard or semihard partons created near the surface are most responsible for the near-side jet, then the away-side jet should experience fully the effect of the medium.  That is indeed the case when one calculates $\langle q' \rangle$ and $\langle k' \rangle$ for various values of $p_t$ and $p_b$ at fixed $\xi$.  It is found in Ref.\ [\refcite{6.6}] that for $\xi = 2.9$, $\langle q' \rangle/\langle k' \rangle$ is in the range of 0.15 to 0.3 for $4 < p_t < 10$ GeV/c and $2 < p_b<  6$ GeV/c, and gives a quantitative measure of the degree of energy loss.  Unfortunately, that cannot be checked experimentally, since only centrality can be selected in realistic collisions, not the dynamical path length $\xi$.

At fixed centrality the value of $\xi$ can fluctuate over a wide range, as can be seen in Fig.\ 11(c).  For $c = 0.05$ the average $\bar{\xi}(c)$ is 2.9, but the characteristic of the away-side jet is dominated by the lower $\xi$ portion of the range.  The consequence is that $Y^{\rm away}_{\pi\pi} (p_t, p_b, c)$ does not depend sensitively on $p_t$, as Fig.\ 14(b) has already indicated for any fixed $c$.  That is in sharp contrast from the case where $\xi$ is fixed at 2.9, and $Y^{\rm away}_{\pi\pi} (p_t, p_b, \xi)$ decreases by an order of magnitude as $p_t$ is decreased from 9 to 4 GeV/c at any fixed $p_b$, as shown in Fig.\ 18(b).  That diminishing yield is because $\langle k \rangle$ is lower at lower $p_t$ and the energy loss by the recoil parton traversing the thick medium results in reduced probability of producing a pion at fixed $p_b$.  That is not the case in Fig.\ 14(b).  At fixed $c$, the decrease of $p_t$ does not lead to significantly lower $Y^{\rm away}_{\pi\pi} (p_t, p_b, c)$ because the hard-scattering point is already in a region of lower $\xi$ to minimize energy loss.  Lowering $p_t$ increases the yield at fixed $p_b$ due to softer parton, but the number of triggers is also higher, so the yield per trigger remain nearly unchanged.  In other words, allowing $\xi$ to be small even at small $c$ removes energy loss as a decisive factor in the problem.  That is what makes dijet tomography ineffective as a probe to learn about a medium that has no fixed thickness.

Further insight can be gained by studying $\langle \beta t'  \rangle$ at fixed $c$, where $t'$ denotes the distance from the hard-scattering point to the away-side surface.  Fig.\ 19(a) shows  $\langle \beta t'  \rangle$ vs $p_b$ for various $p_t$ and $c$.\cite{6.6}  A general impression from that figure is that  $\langle \beta t'  \rangle$ is low, less than approximately 0.4.  That is much lower than $\beta L = 2.9$, which is the average dynamical path length determined from fitting the single-pion distribution from $c = 0.05$ [see Fig.\ 11(b)].  The height of that distribution at large $p_T$ [see Fig.\ 11(a)] is what renders $R_{AA} \approx 0.2$, a number that pQCD calculations aim to obtain.  The fit in Ref.\ [\refcite{6.6}] is achieved by setting $\beta L = 2.9$ in order to obtain the correct normalization for $dN_{\pi}/p_t dp_t$ at large $p_T$, for which the contributions from all partons, near or far, hard or semihard, are counted.  The result that $\langle \beta t'  \rangle \ll \beta L$ shown in Fig.\ 19(a) is therefore indicative of the fact that conditional probability with $p_t$ and $p_b$ fixed is highly restricted compared to the inclusive probability.  With $\langle t'  \rangle$ being much less than $L$ at fixed $c$, one is led to conclude that the unsymmetric dijets are also produced near the surface and are essentially tangential, as with symmetric dijets.  Thus the medium interior is not probed.
This conclusion is distinctively different from the case where $\xi$ is fixed.  Fig.\ 16(b) has shown that the suppression on the away side can be large (hence $\Gamma _{\rm away}$ small) for unsymmetric dijets at $\xi = 2.9$.  In that case $\langle \beta t'  \rangle$ would not be small, as we now have for any fixed $c$.

\begin{figure}[th]
\vspace{-1cm}
\centerline{\psfig{file=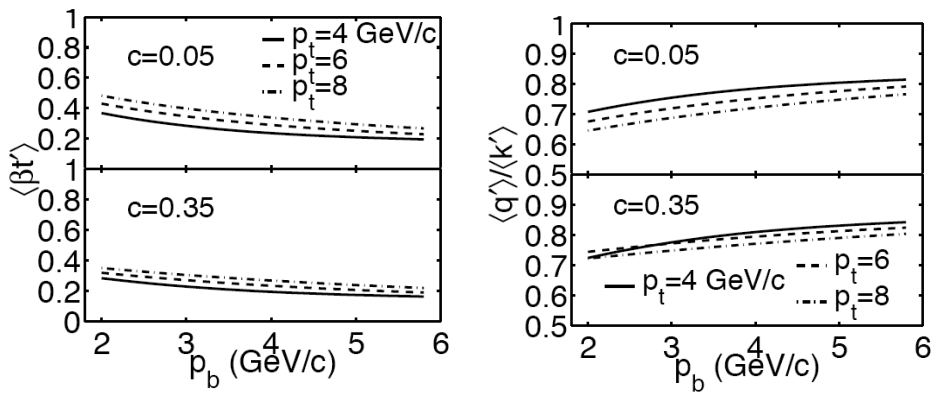,width=14.5cm}}
\vspace*{-5.5cm}
\caption{(a) Left panel: average dynamical path length of recoil parton directed toward the away side. (b) Right panel: Ratio of the average parton momenta of recoil parton at the surface to that at the creation point.}
\end{figure}

Experimental data\cite{6.12} on slightly asymmetric dijets at high $p_T$ show no essential difference in the structures of the two jets and are in support of the findings in Ref.\ [\refcite{6.6}].

Another way to come to the same conclusion is to study the energy loss of the away-side jet.  In Fig.\ 19(b) is shown $\langle q' \rangle/\langle k' \rangle$ for $c = 0.05$ and $0.35$.\cite{6.6}  It is the fraction of average momentum retained by the hard or semihard parton after traversing the medium on the away side.  Being around 0.8 implies that only about 20\% of the parton momentum is lost to the medium, not much more than the fractional energy loss on the near side.  At fixed $p_t$ Fig.\ 19(b) shows an increase of  $\langle q' \rangle/\langle k' \rangle$ with increasing $p_b$ because the hard scattering point is pulled more to the away side, but it shows a decrease of the ratio at increasing $p_t$ since the point is then pulled to the near side.  This push-and-pull effect of $p_t$ and $p_b$ is clearly what one expects in the oppositely-directed jets when the path lengths on the two sides are comparable.  At the symmetry point $p_t = p_b = 4$ GeV/c,  $\langle q' \rangle/\langle k' \rangle$ for $c = 0.05$ is only slightly lower than its value for $c = 0.35$, implying strongly that the fractional energy loss on both sides remains about the same regardless of centrality.  That can only mean that the hard partons are created near the surface and directed tangentially.  Making $p_t \neq p_b$ in unsymmetric dijets does not change $\langle q' \rangle/\langle k' \rangle$ drastically.  Thus so long as the medium thickness cannot be controlled, there seems to be no useful tomography that can be done with parton-initiated dijets. If the away-side jets are dominated by those created near the away-side surface (such as the tangential jets), then the events triggered by direct $\gamma$ on the near side are also likely to be dominated by those where the hard scattering takes place near the away-side surface (not just tangential) so long as an associated particle with significant $p_b$ is required on the away side. The $\pi$-triggered and $\gamma$-triggered distributions, $I_{AA}$, of the associated particles should be roughly the same. There is some experimental evidence for that similarity.\cite{6.13}

It is important to note that the above comment is for tomography only, i.e., medium effect on jets.  The jet effect on medium, such as Mach cone, is a difference matter that depends on different physics and may well reveal properties of the medium interior.

\section{Conclusion}

In this review many problems in heavy-ion collisions have been examined in many parts of the phase space.  The good agreement between theoretical calculation and experimental data in almost all cases cannot but affirm that the successes of the theoretical approach adopted cannot be all fortuitous and that the interpretations given to the physical origins of the measured phenomena are not without some degree of realism.  In some cases there are no apparent alternative schemes to explain the data.  Of course, a phenomenological model is not a theory based on first principles, but its usefulness should not be overlooked when its scope is out of reach by any theory commonly accepted.

In the first part of this review, mainly in Secs.\ 2 and 3, we have summarized the essential basis of the recombination model, answering some questions raised by critics, and pinning down some of the parameters, while pushing the frontier to kinematical regions no other models  have attempted, and establishing parton recombination as a universal hadronization mechanism.   Then in the second and main part recombination is used in an essential way to relate what are observed about jets and ridges to two complementary aspects about the dense medium:  the effects of hard or semihard partons on the medium and the converse.  Those aspects of the medium effects cannot be made empirically relevant unless there is a reliable description of hadronization for all $p_T$ where correlation data exist.

Some of the new insights gained that are of particular current interest are:
\begin{description}
\item[(a)]\quad correlation between trigger and ridge (Sec.\ 4.4),
\item[(b)]\quad independence of $\phi$ anisotropy on fast thermalization (Sec.\ 5.1),
\item[(c)]\quad recombination does not guarantee quark number scaling (Sec.\ 5.3),
\item[(d)]\quad different properties of the near- and away-side jets (Sec.\ 6.3),
\item[(e)]\quad dominance of tangential jets in symmetric and unsymmetric dijets (Secs.\ 6.4 and 6.5).
\end{description}
If there is to be one unifying conclusion to be made as a result of these findings, it is that most observables on jets and ridges are due to hard or semihard partons created near the surface.  Partons created in the interior of the medium that lead to dijets are not able to compete effectively with those that have shorter distance to traverse.  So long as the observables allow the parton creation points to include regions that offer the partons a choice of paths of least resistance, they will take it and dominate.  That is the reality faced by experiments that can only fix centrality, not medium thickness.  With that recognition the efficacy of jet tomography is called into question.  Unlike X-ray scanning of organic or inorganic substances, there is no control of the sources of the hard partons, so the necessary averaging process under-weighs the contribution from the region of the medium that one wants to learn most about.

The above comment refers to parton-initiated dijets.  Of course, for single hadron at large $p_T$ the nuclear modification factor $R_{AA}$ has long been used as a measure of energy loss in dense medium in experiments and in theory.    As soon as a condition is imposed on the detector of another particle on the away side, the region of the system probed in changed.  If that hadron's $p_T$ is low, it may be in the double-humped shoulder region, which does provide some information on what the effect of the away-side jet  is on the medium.  However, to learn about the effect of the main body of the medium on the jet from dihadron correlation is more difficult.

Ridgeology addresses a different set of problems, quite distinct from dijets.  Ridges are stimulated by jets, but are not a part of the jets that are characterized by the participation of shower partons.  Considerable attention has recently been drawn to the study of ridges.  At this point there is no consensus in their theoretical interpretation.  The connection between the ridges found in triggered events\cite{4.7,7.1} and those found in autocorrelation without triggers\cite{4.11,7.2} is in our opinion tenuous, until the $p_T$ dependence of the ridge is clarified.  Minijets\cite{7.3} peaking at $y_t \sim 2.8$ correspond  to $\langle p_T\rangle \sim 1.2$ GeV/c, which is much lower than the $p_T$ range of the triggers used in ridge analysis.  The ridge found in autocorrelation has no $p_T$ cut, while those found in triggered events have $p_T^{\rm assoc}>2$ GeV/c.\cite{4.7}
Nevertheless, there may exist a connection between autocorrelation and the ridges without trigger discussed in Sec.\ 4.2.  

Another phenomenon of some current interest is the observation of extended ridge at large $\Delta\eta$.\cite{7.1} Before one concludes that such a long-range correlation can only arise from the mechanism of strings (or color flux tubes) being stretched between forward- and backward-going quarks, it seems prudent to allow firstly the possibility of other types of early-time dynamics, and secondly 
 a broader view of the problem of particle production at large $|\eta|$. In particular,  one should consider the role played by the hard parton that leads to both the trigger and the ridge. Moreover, one 
 should consider the issue of large $p/\pi$ ratio at $\eta=3.2$ (see Sec.\ 3.4) as a part of the solution of the bigger problem. If the observed large $p/\pi$ ratio is a feature of the final-state interaction (FSI), one should not regard the ridge phenomenon as a manifestation of the initial-state interaction (ISI) only without taking into consideration also the effects of FSI. As we have noted at a number of places throughout this review, hadronization is an important link between the observables and partonic dynamics. The characteristic of ridge formation in azimuthal correlation discussed in Sec.\ 4.4 is a good example of the interplay between ISI and FSI. It would be surprising if the same does not hold true for the correlation in rapidity.

The considerations given in this review to jets and ridges may not be directly relevant at LHC where jets are copiously produced --- unless $p_T$ is extremely high.  At RHIC the background to a rare jet is thermal, but at LHC the background to a high-$p_T$ jet, say at $p_T \sim 100$ GeV/c, includes many other lower-$p_T$ jets.  The admixture of thermal and shower partons in the background introduces new complications to the notion of enhanced thermal partons, and renders what is simply conceived for RHIC inadequate for LHC.  That new energy frontier will indeed open up a wide new horizon.

\section*{Acknowledgment}
I am grateful to many who have helped me in the studies described here, but especially to Charles Chiu and Chunbin Yang.  This work was supported, in part,  by the U.\ S.\
Department of Energy under Grant No. DE-FG02-92ER40972.

\end{document}